\documentclass{pic2012}

\usepackage{atlasphysics}
\usepackage{hypernat}
\usepackage[dvipdfm]{hyperref}
\usepackage{subfigure}
\usepackage{multirow}
\usepackage{heppennames2}
\usepackage{cite}

\def\ifb{\mbox{fb$^{-1}$}}
\def\antibar#1{\ensuremath{#1\bar{#1}}}
\def\bbbar{\antibar{b}}

\newcommand{\ttH}{\ensuremath{q\bar{q}/gg \to t\bar{t}H}}
\newcommand{\hWWenmun}{\ensuremath{H{\rightarrow\,}WW^{(*)}{\rightarrow\,}e\nu\mu\nu}}

\newcommand{\hwwenmun}{\ensuremath{H{\rightarrow\,}WW^{(*)}{\rightarrow\,}e\nu\mu\nu}}
\newcommand{\hWWlnln}{\ensuremath{H{\rightarrow\,}WW^{(*)}{\rightarrow\,}\ell\nu\ell\nu}}
\newcommand{\hwwlnln}{\ensuremath{H{\rightarrow\,}WW^{(*)}{\rightarrow\,}\ell\nu\ell\nu}}
\newcommand\htollll{$H{\rightarrow\,}ZZ^{(*)}{\rightarrow\,}4\ell$}
\newcommand{\hgg}{\ensuremath{H{\rightarrow\,}\gamma\gamma}}
\newcommand{\hZZllll}{\ensuremath{H{\rightarrow\,}ZZ^{(*)}{\rightarrow\,}4\ell}}
\newcommand\htollllp{$H\to ZZ^{(*)}\to 4\ell$}

\newcommand{\mh}{\ensuremath{m_{H}}}
\newcommand{\infb}{fb$^{-1}$}

\newcommand{\Cc}{\ensuremath{\upkappa}}

\newcommand{\masspeak}{126.0}

\newcommand{\massresultStatSys}{\masspeak\,$\pm$\,0.4\,(stat)\,$\pm$\,0.4\,(sys)\,GeV}
\newcommand\excludedrangeaBrief{131--162}
\newcommand\excludedrangebBrief{170--460}
\newcommand\excludedrangeexpaBrief{124--164}
\newcommand\excludedrangeexpbBrief{176--500}
\newcommand{\lowerExpNoGeV}{110}
\newcommand{\lowerExp}{\lowerExpNoGeV\,GeV}
\newcommand{\upperExpNoGeV}{582}
\newcommand{\upperExp}{\upperExpNoGeV\,GeV}
\newcommand{\lowerlowerObsNoGeV}{111}
\newcommand{\upperlowerObsNoGeV}{122} 
\newcommand{\lowerObsNoGeV}{131}
\newcommand{\lowerObs}{\lowerObsNoGeV\,GeV}
\newcommand{\upperObsNoGeV}{559}
\newcommand{\upperlowerObs}{\upperlowerObsNoGeV\,GeV}
\newcommand{\upperObs}{\upperObsNoGeV\,GeV}
\newcommand{\significance}{$6.0\,\sigma$}
\newcommand{\expectedsignificance}{$4.9\,\sigma$}
\newcommand{\significanceESS}{$5.9\,\sigma$}

\newcommand{\ptt}{\ensuremath{p_{\mathrm{Tt}}}}
\newcommand{\Zjets}{$Z$+jets}
\newcommand{\Wjets}{$W$+jets}
\newcommand{\ztoee}{\ensuremath{Z{\rightarrow\,}e^+e^-}}
\newcommand{\lelm}{\ensuremath{H{\rightarrow\,}WW^{(\ast)}{\rightarrow\,}e \nu \mu \nu}}
\newcommand{\metrel}{\ensuremath{E_{\rm T,rel}^{\rm miss}}}
\newcommand{\ZeroJet}{\mbox{0-jet}}
\newcommand{\OneJet}{\mbox{1-jet}}
\newcommand{\TwoJet}{\mbox{2-jet}}

\newcommand{\mT}{\ensuremath{m_{\rm T}}}

\def\CaloMETCutem{\ensuremath{25}}

\def\runauthor{A. Schaffer on behalf of the ATLAS Collaboration}
\def\shorttitle{Higgs Searches in ATLAS}
\begin{document}

\title{Higgs Searches in ATLAS}

\author{A. Schaffer \\
O\lowercase{n behalf of the} ATLAS C\lowercase{ollaboration}}

\address{LAL, Univ Paris-Sud, IN2P3/CNRS\\
Orsay, France\\
E-mail: R.D.Schaffer@cern.ch }

\maketitle

\abstracts{ This talk covers the results of a search for the Standard
  Model Higgs boson in proton-proton collisions with the ATLAS
  detector at the LHC.  The datasets used correspond to integrated
  luminosities of approximately 4.8~\ifb\ collected at
  $\sqrt{s}=7$\,TeV in 2011 and 5.8~\ifb\ at $\sqrt{s}=8$\,TeV in
  2012.  Individual searches in the channels \htollll, \hgg\ and
  $\hWWenmun$ in the 8\,TeV data are combined with previously
  published results of searches for $H{\rightarrow\,}ZZ^{(*)}$,
  $WW^{(*)}$, \bbbar\ and $\tau^+\tau^-$ in the 7\,TeV data and
  results from improved analyses of the \htollll\ and \hgg\ channels
  in the 7\,TeV data.  Clear evidence for the production of a neutral
  boson with a measured mass of \massresultStatSys\ is presented. This
  observation, which has a significance of 5.9 standard deviations,
  corresponding to a background fluctuation probability of $1.7\times
  10^{-9}$, is compatible with the production and decay of the
  Standard Model Higgs boson.  First measurements of the couplings of
  this particle are presented and are compatible with a SM Higgs boson
  hypothesis.  }

\section{Introduction}

The Standard Model (SM) of particle
physics~\cite{np_22_579,prl_19_1264,sm_salam,tHooft:1972fi} has been
tested by many experiments over the last four decades and has been
shown to successfully describe high energy particle
interactions. However, the mechanism that breaks electroweak symmetry
in the SM has not been verified experimentally.  This
mechanism~\cite{Englert:1964et,Higgs:1964ia,Higgs:1964pj,Guralnik:1964eu,Higgs:1966ev,Kibble:1967sv},
which gives mass to massive elementary particles, implies the
existence of a scalar particle, the SM Higgs boson.  The search for
the Higgs boson is one of the highlights of the Large Hadron
Collider~\cite{1748-0221-3-08-S08001} (LHC) physics programme.

Indirect limits on the SM Higgs boson mass of $m_H<158$\,GeV at $95\%$
confidence level (CL) have been set using global fits to precision
electroweak results~\cite{lepew:2010vi}.  Direct searches at
LEP~\cite{Barate:2003sz}, the
Tevatron~\cite{Aaltonen:2012if,D0combo072012,TevatronCombo:2012} and
the LHC~\cite{paper2012prd,Chatrchyan:2012tx} have previously
excluded, at 95\%\ CL, a SM Higgs boson with mass below 600\,GeV,
apart from some mass regions between 116\,GeV and 127\,GeV.

Recently both the ATLAS and CMS Collaborations have reported the
observation of a new particle in the search for the Higgs
boson~\cite{ATLAS:2012af,Chatrchyan:2012ia}.  The CDF and {D\O}
experiments at the Tevatron have also recently reported a broad excess
in the mass region 120--135\,GeV with an observed local significance
for \mH\,=\,125\,GeV of 2.8\,$\sigma$ for the combination of the two
experiments~\cite{TevatronCombo:2012}.  This talk covers the results
reported in~\cite{ATLAS:2012af}, as well as new results on the
couplings of the newly observed particle reported
in~\cite{CouplingsConf}.  More complete discussion of these analyses
is provided in these references.

The data taken are affected by multiple $pp$ collisions occurring in
the same or neighbouring bunch crossings (pile-up). In the 7\,TeV
data, the average number of interactions per bunch crossing was
$\sim\,10$, increasing to $\sim\,20$ in the 8\,TeV data.  The
reconstruction, identification and isolation criteria used for
electrons and photons in the 8\,TeV data are improved, making the
\hZZllll\ and \hgg\ searches more robust against the increased
pile-up. These analyses were re-optimised with simulation and frozen
before looking at the 8\,TeV data.

In the \hWWlnln\ channel, the increased pile-up deteriorates the event
missing transverse momentum, $\MET$, resolution, which results in
significantly larger Drell-Yan background in the same-flavour final
states. Since the $e\mu$ channel provides most of the sensitivity of
the search, only this final state is used in the analysis of the
8\,TeV data.  The kinematic region in which a SM Higgs boson with a
mass between 110\,GeV and 140\,GeV is searched for was kept
blinded during the analysis optimisation, until satisfactory agreement
was found between the observed and predicted numbers of events in
control samples dominated by the principal backgrounds.

\section{The ATLAS detector\label{sec:detector}}

The ATLAS detector~\cite{atlas-det} is a multi-purpose particle
physics detector with approximately forward-backward symmetric
cylindrical geometry.  The inner tracking detector (ID) with a
pseudo-rapidity coverage of $\left| \eta \right|<2.5$ and consists of
a silicon pixel detector, a silicon micro-strip detector, and a
transition radiation tracker.  The ID is surrounded by a thin
superconducting solenoid providing a 2\,T axial magnetic field.  A
high-granularity lead/liquid-argon (LAr) sampling calorimeter measures
the energy and the position of electromagnetic showers within $\left|
  \eta \right|<3.2$.  LAr sampling calorimeters are also used to
measure hadronic showers in the end-cap ($1.5<\left| \eta
\right|<3.2$) and forward ($3.1<\left| \eta \right|<4.9$) regions,
while an iron/scintillator tile calorimeter measures hadronic showers
in the central region ($\left| \eta \right|<1.7$).  The muon
spectrometer (MS) surrounds the calorimeters and consists of three
large superconducting air-core toroid magnets, each with eight coils,
a system of precision tracking chambers ($\left| \eta \right|<2.7$),
and fast tracking chambers for triggering.  A three-level trigger
system selects events to be recorded for offline analysis.

\section{Signal and background simulation samples \label{sec:papersamples}}

The SM Higgs boson production processes considered in this analysis
are the dominant gluon fusion ($gg\to H$, denoted ggF), vector-boson
fusion ($qq'\to qq'H$, denoted VBF) and Higgs-strahlung ($qq'\to WH,
ZH$, denoted $WH$/$ZH$).  The small contribution from the associated
production with a \ttbar\ pair (\ttH, denoted $t\bar{t}H$) is taken
into account only in the \hgg\ analysis. Full details on the
simulation samples is provided in ~\cite{ATLAS:2012af}.

\section{\texorpdfstring{\htollllp{} channel}{H->ZZ(*)->4l channel}}
\label{sec:h4l}

The search for the SM Higgs boson through the decay \htollllp{}, where
$\ell=e$ or $\mu$, provides good sensitivity over a wide mass range
($110$-$600$\,GeV) due to a fully reconstructed final state with
excellent momentum resolution allowing a peak to be seen above the
background. This analysis searches for Higgs boson candidates by
selecting two pairs of isolated leptons, each of which is comprised of
two leptons with the same flavour and opposite charge.  The expected
cross section times branching ratio for the process \htollllp{} with
Higgs mass hypothesis \mH\,=\,125\,GeV is 2.2\,fb for
$\sqrt{s}=7$\,TeV and 2.8\,fb for $\sqrt{s}=8$\,TeV.

The largest background comes from continuum $(Z^{(*)}/\gamma^{*})
(Z^{(*)}/\gamma^{*})$ production, referred to hereafter as
$ZZ^{(*)}$. For low masses there are also important background
contributions from $Z+\rm{jets}$ and $t\bar{t}$ production, where
charged lepton candidates arise either from decays of hadrons with
$b$- or $c$-quark content or from mis-identification of jets. The
reducible backgrounds are suppressed with isolation and impact
parameter requirements. 

The crucial experimental aspects of this channel are:
\begin{itemize}
\item high lepton acceptance, reconstruction and identification down
  to low $\pt$,
\item good lepton energy/momentum resolution,
\item good control of reducible backgrounds ($Z+\rm{jets}$ and
  $Z+b\bar{b}$ and $t\bar{t}$) in low mass region below
  $\sim$\,170\,GeV.
  
  The reducible background estimates cannot rely on Monte Carlo (MC)
  simulation alone due to theoretical uncertainties and $b/jet \to
  \ell$ modeling. So the MC simulation is validated with data enriched
  control samples.
\end{itemize}

The 7\,TeV data have been re-analysed and combined with the 8\,TeV 
data. The analysis is improved in several aspects with respect
to~\cite{ATLAS:2012ac} to enhance the sensitivity to a low-mass Higgs
boson. In particular, the kinematic selections are revised.  The
8\,TeV data analysis also benefits from improvements in the electron
reconstruction and identification. The expected local signal
significances for a Higgs boson with \mH\,=\,125\,GeV are
$1.6\,\sigma$ for the 7\,TeV data (to be compared with
$1.25\,\sigma$ in~\cite{ATLAS:2012ac}) and $2.1\,\sigma$ for the
8\,TeV data.
  
\subsection{Event selection} 

The data are selected using single-lepton or di-lepton triggers.  Muon
candidates are formed by matching reconstructed ID tracks with either
a complete track or a track-segment reconstructed in the
MS~\cite{ATLAS-CONF-2011-063}.  Electron candidates are formed from ID
tracks pointing to electromagnetic calorimeter clusters, where the
cluster must satisfy a set of identification
criteria~\cite{Aad:2011mk} that require the longitudinal and
transverse shower profiles to be consistent with those expected for
electromagnetic showers. The electron tracks are fitted using a
Gaussian-Sum Filter~\cite{GSFConf}, allowing for bremsstrahlung energy
losses.

Quadruplets are formed from same-flavour opposite-charge (SFOC) lepton
pairs with their transverse momentum, $\pt$, at least 20,15,10\,GeV
for the three leading leptons, and at least 7(6)\,GeV for the final
muon (electron).  The leading SFOC lepton pair has an invariant mass
($m_{12}$) closest to the $Z$ boson mass with 50\,GeV\,$< m_{12}
<$\,106\,GeV.  The sub-leading SFOC lepton pair, with its invariant
mass ($m_{34}$) required to be in the range
$m_{\rm{min}}<m_{34}<115$\,GeV with $m_{\rm{min}}$ varying from
17.5\,GeV at $m_{4\ell}$\,=\,120\,GeV to 50\,GeV at
$m_{4\ell}$\,=\,190\,GeV~\cite{ATLAS-CONF-2012-092}. There are four
different analysis sub-channels: $4e$, $2e2\mu$, $2\mu2e$ and $4\mu$,
with the cross-flavoured pairs ordered in $\pt$.

Non-prompt leptons from heavy flavour decays, electrons from photon
conversions and jets mis-identified as electrons have broader
transverse impact parameter distributions than prompt leptons from
$\Zboson$ boson decays and/or are non-isolated. Thus, the \Zjets\ and
$t\bar{t}$~background contributions are reduced by applying a cut on
the transverse impact parameter significance, $d_0/\sigma_{d_0}$,
required to be less than 3.5 (6.5) for muons (electrons).  In
addition, leptons must satisfy isolation requirements based on
tracking and calorimetric information.  The normalised track isolation
discriminant requires the sum of the transverse momenta of tracks
inside a cone of size $\Delta R=0.2$ around the lepton direction
divided by the lepton $\pt$ be smaller than 0.15.  The normalised
calorimetric isolation for electrons, the sum of the $\et$ of
topological clusters~\cite{Lampl:1099735} within $\Delta R=0.2$
divided by the electron $\et$, is required to be less than 0.2.
Finally, the normalised calorimetric isolation discriminant for muons,
the \et~sum of the calorimeter cells inside $\Delta R=0.2$ divided by
the muon $\pt$, is required to be less than 0.3.

The combined signal reconstruction and selection efficiencies for a SM
Higgs with \mH\,=\,125\,GeV for the 7\,TeV (8\,TeV) data are $37$\%
($36$\%) for the $4\mu$ channel, $20$\% ($22$\%) for the
$2e2\mu$/$2\mu2e$ channels and $15$\% ($20$\%) for the $4e$ channel.

The $4\ell$ invariant mass resolution is improved by applying a
$Z$-mass constrained kinematic fit to the leading lepton pair for
$m_{4\ell}<190$\,GeV and to both lepton pairs for higher masses. The
expected width of the reconstructed mass distribution is dominated by
the experimental resolution for $m_{H}<350$\,GeV, and by the natural
width of the Higgs boson for higher masses ($30$\,GeV at
\mH\,=\,400\,GeV).  The typical mass resolutions for \mH\,=\,125\,GeV
are $1.8$\,GeV, $2.0$\,GeV and $2.5$\,GeV for the $4\mu$,
$2e2\mu$/$2\mu2e$ and $4e$ sub-channels, respectively.

\subsection{Background estimation}

The expected background yield and composition are estimated using the
MC simulation normalised to the theoretical cross section for
$ZZ^{(*)}$ production and by methods using control regions from data
for the $Z+\rm{jets}$ and $t\bar{t}$ processes. Since the background
composition depends on the flavour of the sub-leading lepton pair,
different approaches are taken for the $\ell\ell+\mu\mu$ and the
$\ell\ell+ee$ final states. The transfer factors needed to extrapolate
the background yields from the control regions defined below to the
signal region are obtained from the MC simulation. The MC description
of the selection efficiencies for the different background components
has been verified with data.

The reducible $\ell\ell+\mu\mu$ background is dominated by $t\bar{t}$
and $Z+{\rm jets}$ (mostly $Z+b\bar{b}$) events. A control region is
defined by removing the isolation requirement on the leptons in the
sub-leading pair, and by requiring that at least one of the
sub-leading muons fails the transverse impact parameter significance
selection. These modifications remove $ZZ^{(*)}$ contributions, and
allow both the $t\bar{t}$ and $Z+{\rm jets}$ backgrounds to be
estimated simultaneously using a fit to the $m_{12}$ distribution
($m_{12}$ peaks at $m_{\Zboson}$ for $Z+{\rm jets}$ and $t\bar{t}$ is
relatively flat in $m_{12}$).

In order to estimate the reducible $\ell\ell+ee$ background, a control
region is formed by relaxing the selection criteria for the electrons
of the sub-leading pair. The different sources of electron background
are then separated into categories consisting of non-prompt leptons
from heavy flavour decays, electrons from photon conversions and jets
mis-identified as electrons, using appropriate discriminating
variables~\cite{Aad:2011rr}.  This method allows the sum of the
$Z+{\rm jets}$ and $t\bar{t}$ background contributions to be
estimated.  Two other methods have been used as cross-check and yield
consistent results.

\begin{table}[!htb]
  \centering
  \caption{Summary of the estimated numbers of $Z+{\rm jets}$ and
    $t\bar{t}$ background events,  for the 7\,TeV and 8\,TeV
    data in the entire phase-space of the analysis after the kinematic
    selections described in the text. The backgrounds are combined for
    the $2\mu2e$ and $4e$ channels, as discussed in the text. The
    first uncertainty is statistical, while the second is
    systematic. Ref.~\protect\cite{ATLAS:2012af}. \label{tab:bkg_overview}} 
  \vspace{0.2cm}
  \scalebox{0.9}{
    \begin{tabular}{cr@{$\pm$}c@{$\pm$}lr@{$\pm$}c@{$\pm$}l}
      \hline\hline
      Background& \multicolumn{6}{c}{Estimated}  \\
      & \multicolumn{6}{c}{numbers of events}  \\
      & \multicolumn{3}{c}{$\sqrt{s}=7$\,TeV}&\multicolumn{3}{c}{$\sqrt{s}=8$\,TeV}  \\
      \hline
      &\multicolumn{6}{c}{$4\mu$}\\ 
      \hline
      \Zjets               & 0.3&  0.1&  0.1& 0.5  &  0.1  &  0.2\\  
      $t\bar{t}$           &0.02&  0.02& 0.01& 0.04 &  0.02 &  0.02\\
      \hline
      &\multicolumn{6}{c}{$2e2\mu$}\\ 
      \hline
      \Zjets              &0.2  &  0.1   &  0.1  & 0.4  &   0.1  &  0.1\\
      $t\bar{t}$          &0.02  &  0.01  &  0.01 & 0.04 &  0.01  &  0.01\\
      \hline
      &\multicolumn{6}{c}{$2\mu2e$}\\ 
      \hline
      \Zjets, $t\bar{t}$     &2.6 & 0.4 & 0.4 & 4.9  &  0.8  &  0.7 \\
      \hline
      &\multicolumn{6}{c}{$4e$}\\ 
      \hline
      \Zjets, $t\bar{t}$     &3.1 & 0.6 & 0.5  & 3.9  &  0.7  &  0.8\\
      \hline
      \hline
    \end{tabular}}
\end{table}

\begin{figure}[!htb]
  \centering 
  \includegraphics[width=0.60\textwidth]{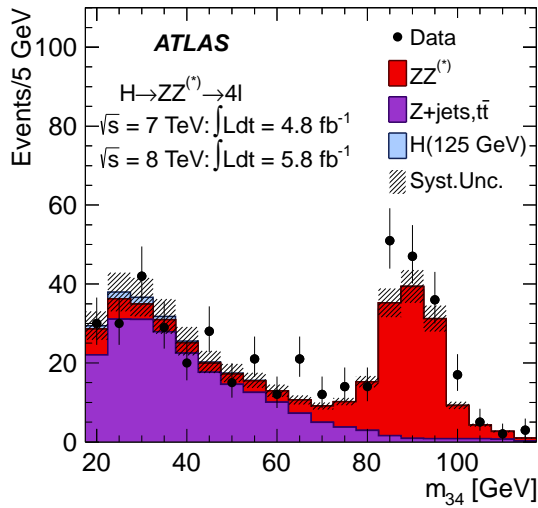}
  \caption{Invariant mass distribution of the sub-leading lepton pair
    ($m_{34}$) for a sample with of a $Z$ boson candidate and an
    additional same-flavour electron or muon pair, for the combined
    7\,TeV and 8\,TeV data after the kinematic selections described in
    the text.  Isolation and transverse impact parameter significance
    requirements are applied to the leading lepton pair only. The MC
    is normalised to the data-driven background estimations. The
    relatively small contribution of a SM Higgs with \mH\,=\,125\,GeV
    in this sample is also shown.
    Ref.~\protect\cite{ATLAS:2012af}.\label{fig:sub_cr_all}} 
\end{figure}

The data-driven background estimates are summarised
in Table~\ref{tab:bkg_overview}. The distribution of $m_{34}$, for
events selected by the analysis except that the isolation and
transverse impact parameter requirements for the sub-leading lepton pair are removed, is
presented in Fig.~\ref{fig:sub_cr_all}. 

\subsection{Systematic uncertainties}\label{sec:h4lsyst}

The uncertainties on the integrated luminosities are determined to be
1.8\%\ for the 7\,TeV data and 3.6\%\ for the 8\,TeV data using
the techniques described in~\cite{ATLAS-CONF-2012-080}.

The uncertainties on the lepton reconstruction and identification
efficiencies and on the momentum scale and resolution are determined
using samples of $W$, $Z$ and $J/\psi$
decays~\cite{ATLAS-CONF-2011-063,Aad:2011mk}.  The relative
uncertainty on the signal acceptance due to the uncertainty on the
muon reconstruction and identification efficiency is $\pm 0.7\%$ ($\pm
0.5\%$/$\pm 0.5\%$) for the $4\mu$ ($2e2\mu$/$2\mu2e$) channel for
$m_{4\ell}$\,=\,600\,GeV and increases to $\pm 0.9\%$ ($\pm
0.8\%$/$\pm 0.5\%$) for $m_{4\ell}$\,=\,115\,GeV. Similarly, the
relative uncertainty on the signal acceptance due to the uncertainty
on the electron reconstruction and identification efficiency is $\pm
2.6\%$ ($\pm 1.7\%$/$\pm 1.8\%$) for the $4e$ ($2e2\mu$/$2\mu2e$)
channel for $m_{4\ell}$\,=\,600\,GeV and reaches $\pm 8.0\%$ ($\pm
2.3\%$/$\pm 7.6\%$) for $m_{4\ell}$\,=\,115\,GeV.  The uncertainty on
the electron energy scale results in an uncertainty of $\pm 0.7\%$
($\pm 0.5\%$/$\pm 0.2\%$) on the mass scale of the $m_{4\ell}$
distribution for the $4e$ ($2e2\mu$/$2\mu2e$) channel.  The impact of
the uncertainties on the electron energy resolution and on the muon
momentum resolution and scale are found to be negligible.

The theoretical uncertainties associated with the signal are described
in detail in~\cite{ATLAS:2012af}.

\begin{figure}[h!tp]
  \centering
  \subfigure[\label{fig:finalMassesSignal}]{\includegraphics[width=0.49\linewidth]{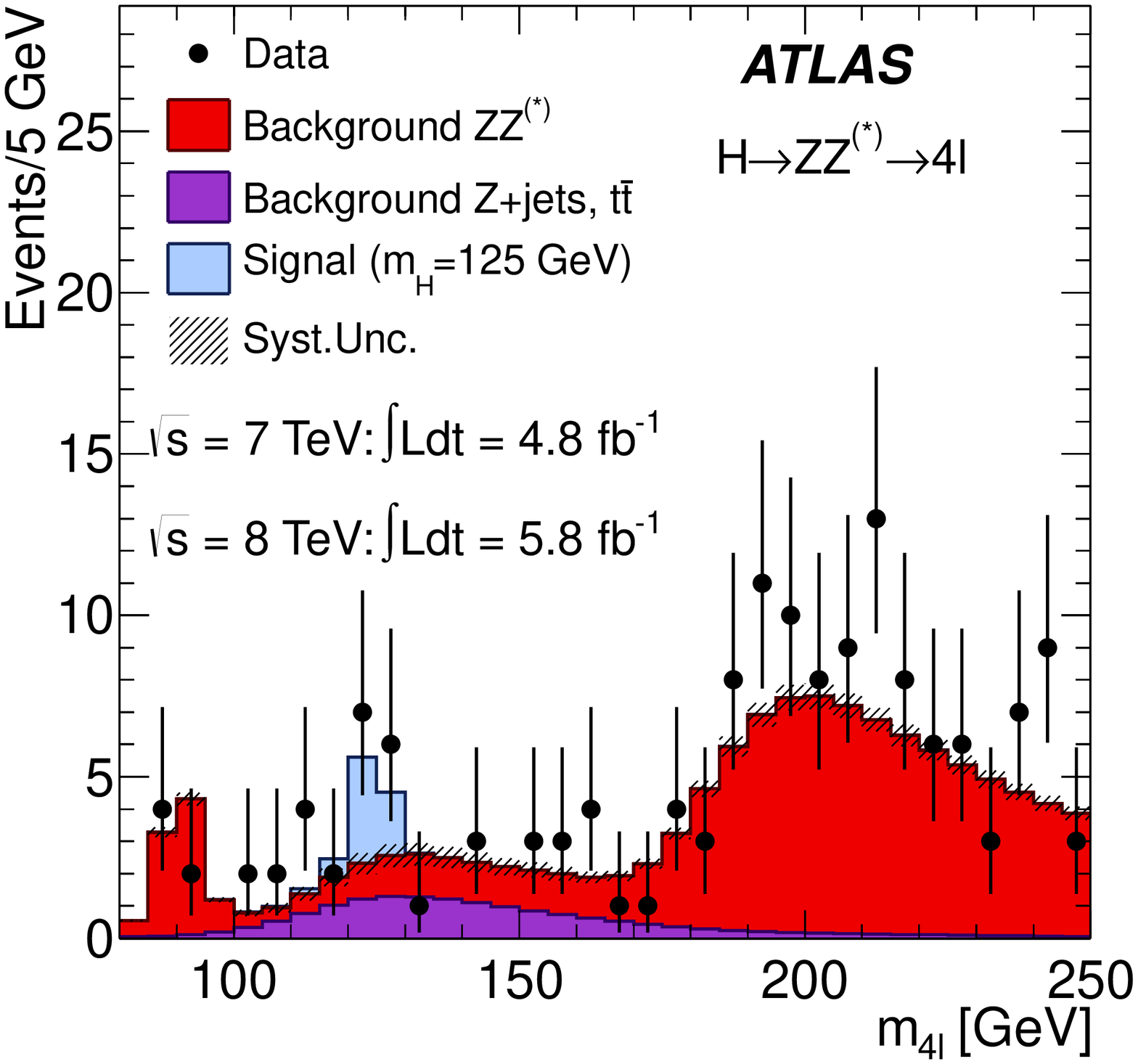}}
  \subfigure[\label{fig:M12vsM34}]{\includegraphics[width=0.49\linewidth]{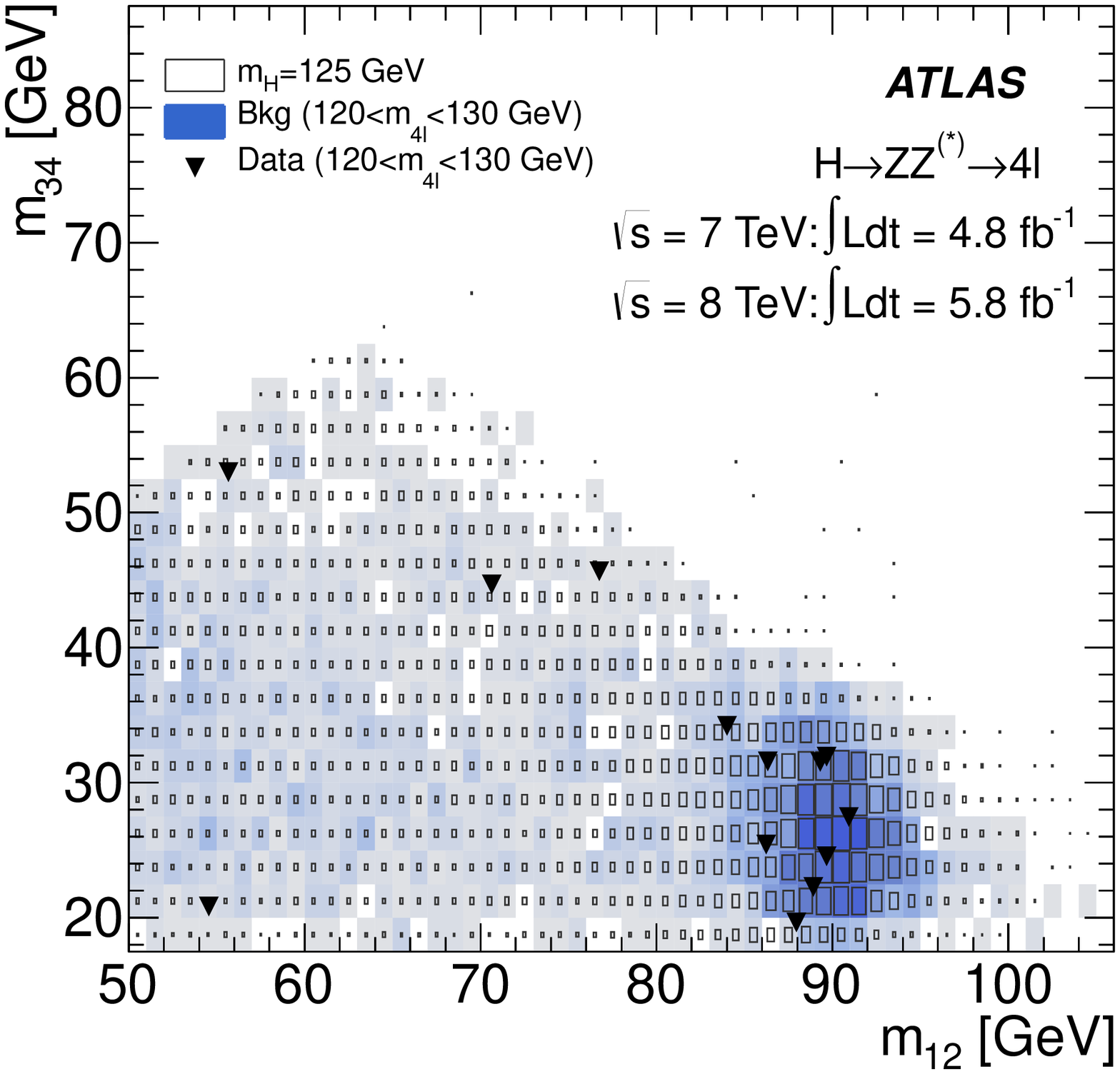}}
  \caption{\subref{fig:finalMassesSignal} The distribution of the
    four-lepton invariant mass, $m_{4\ell}$, for the selected
    candidates, compared to the background expectation in the
    $80$--$250$\,GeV mass range, for the combined 7\,TeV and
    8\,TeV data.  The signal expectation for a SM Higgs with
    \mH\,=\,125\,GeV is also shown. \subref{fig:M12vsM34} The
    distribution of the $m_{34}$ versus the $m_{12}$ invariant mass,
    before the application of the $Z$-mass constrained kinematic fit,
    for the selected candidates in the $m_{4\ell}$ range
    $120$--$130$\,GeV.  The expected distributions for a SM Higgs
    with \mH\,=\,125\,GeV (the sizes of the boxes indicate the
    relative density) and for the total background (the intensity of
    the shading indicates the relative density) are also
    shown. Ref.~\protect\cite{ATLAS:2012af}. \label{fig:finalMassesM4l_M12vsM34}}
\end{figure} 

\subsection{Results}

The expected distributions of $m_{4\ell}$ for the background and for a
Higgs boson signal with \mH\,=\,125\,GeV are compared to the data in
Fig.~\ref{fig:finalMassesSignal}.  The numbers of observed and
expected events in a window of $\pm 5$\,GeV around \mH\,=\,125\,GeV
are presented for the combined 7\,TeV and 8\,TeV data in
Table~\ref{tab:yields}. The distribution of the $m_{34}$ versus
$m_{12}$ invariant mass is shown in Fig.~\ref{fig:M12vsM34}. The
statistical interpretation of the excess of events near
$m_{4\ell}$\,=\,125\,GeV in Fig.~\ref{fig:finalMassesSignal} is
presented in Section~\ref{sec:Results}.

\label{sec:h4lresults}
\begin{table}[h!tb]
  \centering
  \caption{The numbers of expected signal (\mH\,=\,125\,GeV) and
    background events, together with the numbers of observed events in
    the data, in a window of size $\pm 5$\,GeV  around $125$\,GeV,
    for the combined 7\,TeV and 8\,TeV 
    data. Ref.~\protect\cite{ATLAS:2012af}. \label{tab:yields}} 
  \vspace{0.3cm}
  \scalebox{0.85}{
    \begin{tabular}{@{}ccccc@{}}
      \hline\hline
      & Signal & $ZZ^{(*)}$ & $Z+\rm{jets}$,~$t\bar{t}$ & Observed \\
      \hline
      $4\mu$ & 2.09$\pm$0.30 & 1.12$\pm$0.05 & 0.13$\pm$0.04 & 6\\
      \hline
      $2e2\mu$/$2\mu2e$ &2.29$\pm$ 0.33 & 0.80$\pm$0.05 & 1.27$\pm$0.19 & 5\\
      \hline
      $4e$ & 0.90$\pm$0.14 & 0.44$\pm$0.04 & 1.09$\pm$0.20 & 2\\
      \hline\hline
    \end{tabular}}
\end{table}	

\section{\texorpdfstring{\hgg\ channel}{H->gg channel}}
\label{sec:hgg}
The search for the SM Higgs boson through the decay \hgg\ is performed
in the mass range between 110\,GeV and 150\,GeV.  The dominant
background is SM di-photon production~($\gamma\gamma$); contributions
also come from $\gamma$+jet and jet+jet production with one or two
jets mis-identified as photons~($\gamma j$ and $jj$) and from the
Drell-Yan process with the electrons mis-identified as photons.  The
7\,TeV data have been re-analysed and the results combined with
those from the 8\,TeV data.  Among other changes to the analysis, a
new category of events with two forward jets is introduced, which
enhances the sensitivity to the VBF process.  Overall, the sensitivity
of the analysis has been improved by about $20\%$ with respect to that
described in~\cite{ATLAS:2012ad}.

\subsection{Event selection}
\label{sec:hggselect}
The data used in this channel are selected using a di-photon
trigger~\cite{Aad:2012xs}, with $>\,99\%$ efficiency after the final
event selection. Events are required to contain at least one
reconstructed vertex with at least two tracks, as well as two photon
candidates.  Photon candidates must be in the fiducial region
$\left|\eta\right|<2.37$, excluding the calorimeter transition region
$1.37\leq\left|\eta\right|<1.52$.  Converted photons have one or two
tracks matching the clusters in the calorimeter.  The photon
reconstruction efficiency is about $97\%$ for
$E_{\mathrm{T}}>\,30$\,GeV.

MC simulation is used to calibrate for energy losses upstream of the
calorimeter and leakage outside of the cluster of the photon
candidates; this is done separately for unconverted and converted
candidates.  The calibration is refined by applying $\eta$-dependent
correction factors ($\pm 1\%$) determined from measured \ztoee\
events.  The leading (sub-leading) photon candidate is required to
have $E_{\mathrm{T}}>\,40$\,GeV (30\,GeV).

Photon candidates must pass identification criteria based on shower
shapes and hadronic energy leakage~\cite{Aad:2010sp}.  For the
7\,TeV data, the selection uses a neural network, and for the
8\,TeV data, cut-based criteria are used.  This cut-based selection
has been tuned to be robust against pile-up.  The photon
identification efficiencies range from $85\%$ to above $95\%$. To
further suppress the jet background, an isolation selection is applied
by requiring the transverse energy of topological clusters within
$\Delta R < 0.4$ to be less than 4\,GeV.

\subsection{Invariant mass reconstruction}
The invariant mass of the two photons is evaluated using the photon
energies measured in the calorimeter, the azimuthal angle $\phi$
between the photons as determined from the positions of the photons in
the calorimeter, and the values of $\eta$ calculated from the position
of the identified primary vertex and the impact points of the photons
in the calorimeter.

The primary vertex is identified by combining the following
information in a global likelihood: the directions of flight of the
photons as determined using the longitudinal segmentation of the
electromagnetic calorimeter~(calorimeter pointing), the parameters of
the beam spot, and the $\sum\pt^2$ of the tracks associated with each
reconstructed vertex.  In addition, for the 7\,TeV data analysis,
the photon conversion vertex is used in the likelihood.  The
calorimeter pointing is sufficient to ensure that the contribution of
the opening angle between the photons to the mass resolution is
negligible.  The tracking information from the ID improves the
identification of the primary vertex, which is needed for the jet
selection in the 2-jet category.

The number of selected di-photon candidates with an invariant mass
between 100\,GeV and 160\,GeV is 23788 (35251) in the 7\,TeV
(8\,TeV) data sample.

\subsection{Event categorisation}
\label{sec:hggeventcat}
To increase the sensitivity to a Higgs boson signal, the events are
separated into ten mutually exclusive categories having different mass
resolutions and signal-to-background ratios.  An exclusive category of
events containing two jets improves the sensitivity to VBF.  The other
nine categories are defined by the presence or not of converted
photons, $\eta$ of the selected photons, and \ptt, the
component\footnote{$p_{\mathrm{Tt}} = {\left|({{\bf
          p}_\mathrm{T}^{\gamma_1}} + {{\bf p}_\mathrm{T}^{\gamma_2}})
      \times ({{\bf p}_\mathrm{T}^{\gamma_1}} -{{\bf
          p}_\mathrm{T}^{\gamma_2}})\right|/\left|{{\bf
          p}_\mathrm{T}^{\gamma_1}} - {{\bf
          p}_\mathrm{T}^{\gamma_2}}\right|}$, where ${{\bf
      p}_\mathrm{T}^{\gamma_1}}$ and ${{\bf p}_\mathrm{T}^{\gamma_2}}$
  are the transverse momenta of the two photons.}  of the di-photon
\pt\ that is orthogonal to the axis defined by the difference between
the two photon momenta~\cite{PTT_OPAL,PTT_ZBoson}.

\begin{table}[t!]
  \begin{center}
    \caption{Number of events in the data~($N_\mathrm{D}$) and
      expected number of signal events~($N_\mathrm{S}$) for
      \mH\,=\,126.5\,GeV from the \hgg\ analysis, for each category in
      the mass range $100{\--}160$\,GeV.  The mass resolution FWHM
      (see text) is also given for the 8\,TeV data.  The Higgs boson
      production cross section multiplied by the branching ratio into
      two photons~($\sigma\times B(H\to\gamma\gamma)$) is listed for
      \mH\,=\,126.5\,GeV.  The statistical uncertainties on
      $N_\mathrm{S}$ and FWHM are less than
      1\,\%. Ref.~\protect\cite{ATLAS:2012af}. }
    \label{tab:SBnumbers}
    \vspace{0.3cm}
    \scalebox{0.75}{ 
      \begin{tabular}{l|rr|rr|r}
        \hline\hline
        $\sqrt{s}$ & \multicolumn{2}{c|}{7\,TeV} & \multicolumn{3}{c}{8\,TeV} \\
        \hline
        $\sigma\times B(H\to\gamma\gamma)$ [fb]      &       &  39 &       &  50 & FWHM \\
        \cline{1-5}
        Category & $N_\mathrm{D}$ & $N_\mathrm{S}$ & $N_\mathrm{D}$ & $N_\mathrm{S}$ & [GeV] \\
        \hline
        Unconv. central, low \ptt   &  2054 &  10.5 &  2945 &  14.2 & 3.4 \\
        Unconv. central, high \ptt  &    97 &   1.5 &   173 &   2.5 & 3.2 \\
        Unconv. rest, low \ptt      &  7129 &  21.6 & 12136 &  30.9 & 3.7 \\
        Unconv. rest, high \ptt     &   444 &   2.8 &   785 &   5.2 & 3.6 \\
        Conv. central, low \ptt     &  1493 &   6.7 &  2015 &   8.9 & 3.9 \\
        Conv. central, high \ptt    &    77 &   1.0 &   113 &   1.6 & 3.5 \\
        Conv. rest, low \ptt        &  8313 &  21.1 & 11099 &  26.9 & 4.5 \\
        Conv. rest, high \ptt       &   501 &   2.7 &   706 &   4.5 & 3.9 \\
        Conv. transition            &  3591 &   9.5 &  5140 &  12.8 & 6.1 \\
        2-jet                       &    89 &   2.2 &   139 &   3.0 & 3.7 \\
        \hline
        All categories~(inclusive)  & 23788 &  79.6 & 35251 & 110.5 & 3.9 \\
        \hline\hline
      \end{tabular}
    }
  \end{center}
\end{table}

\subsection{Signal modelling}
The description of the Higgs boson signal is obtained from MC
simulation.  For both the 7\,TeV and 8\,TeV MC samples, the fractions
of ggF, VBF, $WH$, $ZH$ and $t\bar{t}H$ production are approximately
$88\%$, $7\%$, $3\%$, $2\%$ and $0.5\%$, respectively, for
\mH\,=\,126.5\,GeV.  The simulated shower shape distributions are
shifted slightly to improve the agreement with the
data~\cite{Aad:2010sp}, and the photon energy resolution is
broadened~(by approximately $1\%$ ($1.2{\--}2.1\%$ ) in the barrel
(end-cap) calorimeter) to account for small differences observed
between \ztoee\ data and MC events.  The signal yields expected for
the 7\,TeV and 8\,TeV data samples are given in
Table~\ref{tab:SBnumbers}.  The overall selection efficiency is about
$40\%$.

The shape of the invariant mass of the signal in each category is
modelled by the sum of a Crystal Ball function~\cite{CSB}, describing
the core of the distribution with a width $\sigma_{CB}$, and a
Gaussian contribution describing the tails~(amounting to $<$$10\%$) of
the mass distribution.  The expected full-width-at-half-maximum (FWHM)
is 3.9\,GeV and $\sigma_{CB}$ is 1.6\,GeV for the inclusive sample, and
varies with event category (see Table~\ref{tab:SBnumbers}).

\subsection{Background modelling}
The background in each category is estimated from data by fitting the
di-photon mass spectrum in the mass range $100{\--}160$\,GeV with a
selected model with free parameters of shape and normalisation.
Different models are chosen for the different categories to achieve a
good compromise between limiting the size of a potential bias while
retaining good statistical power.  The models are a fourth-order
Bernstein polynomial function~\cite{Bernstein}, an exponential
function of a second-order polynomial and an exponential function.
Based on MC studies, the background model which has the best
sensitivity for \mH\,=\,125\,GeV and a bias less of than a $20\%$ of
the statistical uncertainty is chosen for each category.  The largest
absolute signal yield as defined above is taken as the systematic
uncertainty on the background model, amounting to $\pm(0.2{\--}4.6)$
and $\pm(0.3{\--}6.8)$ events, depending on the category for the
7\,TeV and 8\,TeV data samples, respectively.

\subsection{Systematic uncertainties}
The dominant experimental uncertainty on the signal yield~$\pm8\%$
($\pm11\%$) for 7\,TeV  (8\,TeV) data comes from the photon
reconstruction and identification efficiency, which is estimated with
data using electrons from $Z$ decays and photons from
$Z\to\ell^+\ell^-\gamma$ events.  The total uncertainty on the mass
resolution is $\pm14\%$.  The dominant contribution ($\pm12\%$) comes
from the uncertainty on the energy resolution of the calorimeter,
which is determined from \ztoee\ events.

\begin{figure}[h!]
  \begin{center}
    \subfigure[\label{fig:mgg}]{\includegraphics[width=0.49\linewidth]{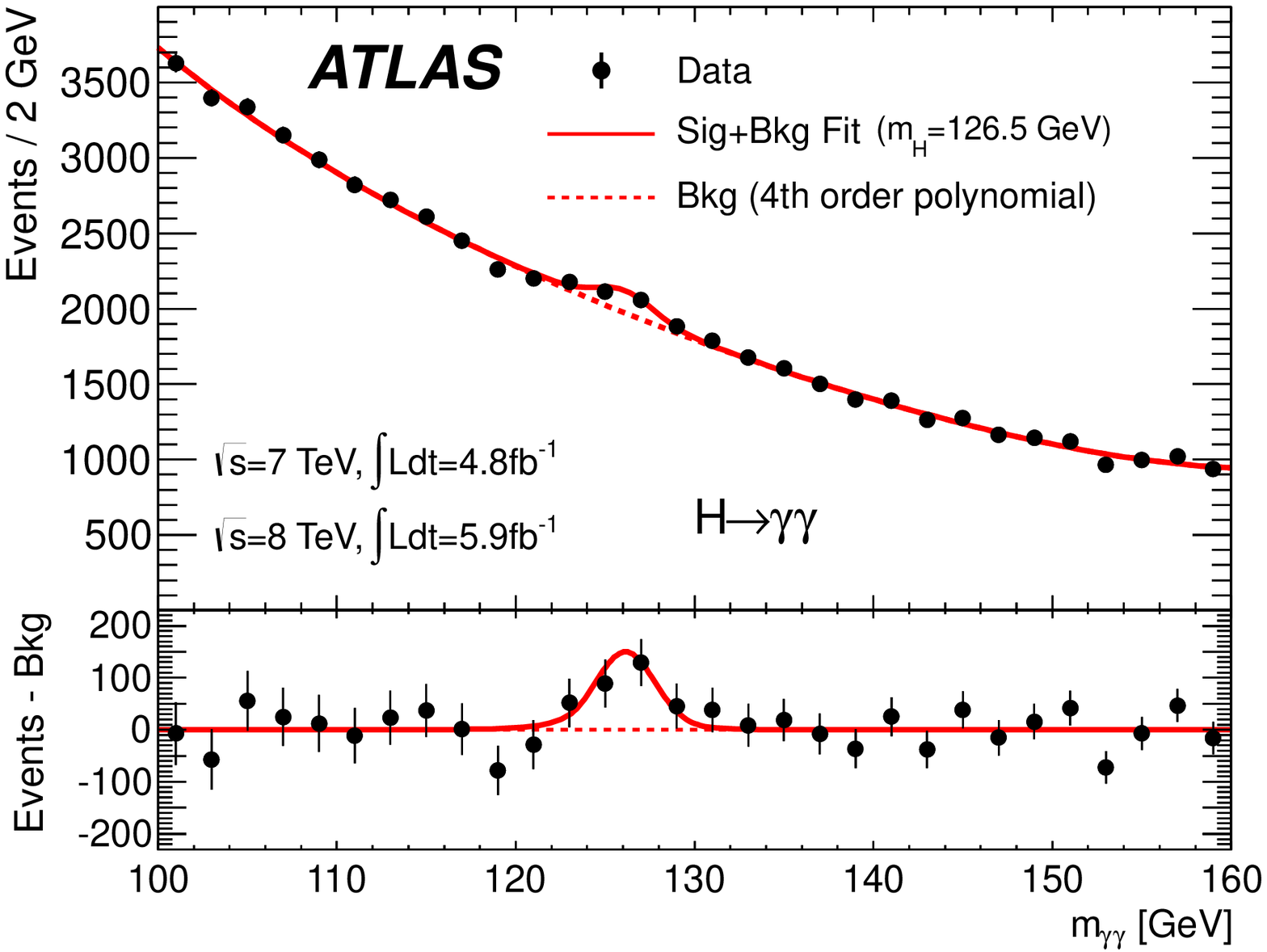}}
    \subfigure[\label{fig:mggweighted}]{\includegraphics[width=0.49\linewidth]{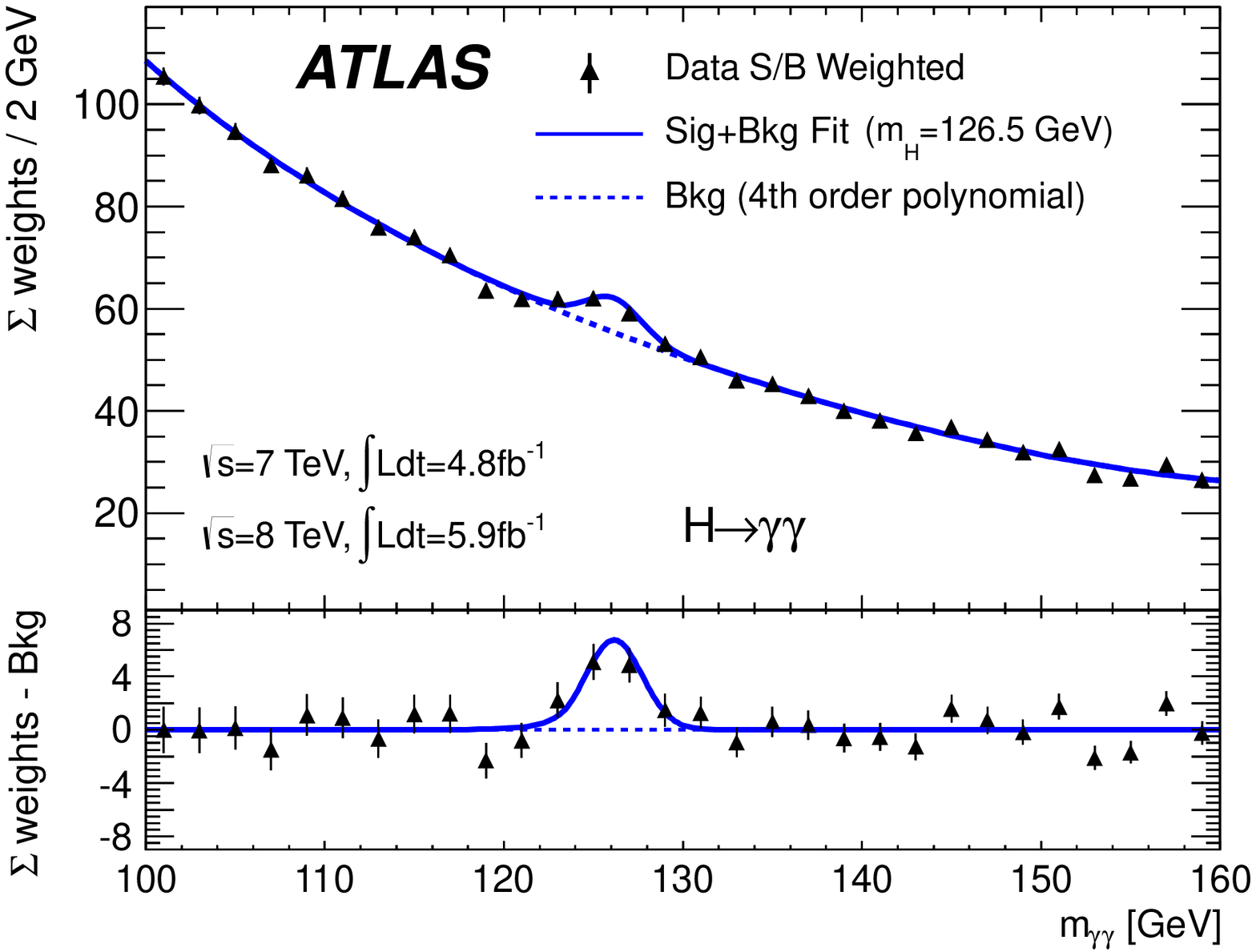}}
  \end{center}
  \caption{The distributions of the invariant mass of di-photon
    candidates after all selections for the combined 7 TeV and 8 TeV
    data sample.  The inclusive sample is shown in \subref{fig:mgg}
    and a weighted version of the same sample in
    \subref{fig:mggweighted}; the weights are explained in the text.
    The result of a fit to the data of the sum of a signal component
    fixed to \mH\,=\,126.5\,GeV and a background component described
    by a fourth-order Bernstein polynomial is superimposed.  The
    residuals of the data and weighted data with respect to the
    respective fitted background component are also
    displayed. Ref.~\protect\cite{ATLAS:2012af}. }
  \label{fig:fitincl}
\end{figure}

\subsection{Results}
\label{sec:hgg:results}
The distribution of the invariant mass, $m_{\gamma\gamma}$, of the
di-photon events, summed over all categories, is shown in
Fig.~\ref{fig:mgg}.  The result of a fit including a signal component
fixed to \mH\,=\,126.5\,GeV and a background component described by a
fourth-order Bernstein polynomial is superimposed.

The statistical analysis uses an unbinned likelihood function
constructed from those of the ten categories of the 7\,TeV and 8\,TeV
data samples.  To demonstrate the sensitivity of this likelihood
analysis, Fig.~\ref{fig:mggweighted} also shows the mass spectrum
obtained after weighting events with category-dependent factors
reflecting the signal-to-background ratios.  The statistical
interpretation of the excess of events near
$m_{\gamma\gamma}$\,=\,126.5\,GeV in Fig.~\ref{fig:fitincl} is
presented in Section~\ref{sec:Results}.

\section{\texorpdfstring{$\hwwenmun$ channel}{H->WW(*)->evmuv channel}}
\label{sec:hww}

The signature for the \hWWlnln\ channel is two opposite-charge leptons
with large transverse momentum and a large momentum imbalance in the
event due to the escaping neutrinos.  This channel has a high rate,
but limited mass resolution. The dominant backgrounds are non-resonant
$WW$ and top production, both of which have real $W$~pairs in the
final state.  Other important backgrounds include Drell-Yan events
($Z/\gamma^{(\ast)}{\to\,}\ell\ell$) with mis-measured \met{}, \Wjets\
events with a fake second lepton, and $W\gamma$ events with an
electron from a conversion.

The analysis of the 8 TeV data is restricted to the $e\mu$ final
state, providing $>\,85\%$ of the sensitivity of the search, due to
the higher luminosity and increased number of interactions which
worsens the Drell-Yan background.  The Drell-Yan background to the
$e\mu$ final state is from semi-lepton $\tau$ decays and thus
significantly reduced.

\subsection{Event selection}
\label{sec:selection}

For the 8 TeV \lelm\ search, the data are selected using inclusive
single-muon and single-electron triggers.  Candidates are selected
with a leading (sub-leading) lepton \et\,$>\,25$\,GeV
($>\,15$\,GeV).  The lepton selection and isolation have more
stringent requirements than those used for the \htollllp{} analysis
(see Section~\ref{sec:h4l}), to reduce the larger background from
non-prompt leptons in the $\ell\nu\ell\nu$ final state.  Events are
separated into 0, 1 and 2-jet categories, with $\pt^{\rm jet} >\,
25(30)$\,GeV for $\left|\eta\right|<\,2.5~(2.5-4.5)$.

With two neutrinos in the signal final state, events are required to
have large \met{}. The quantity $\metrel$ is required to be
$>\,\CaloMETCutem$\,GeV and is the projection of the direction of
\met{} perpendicular to the nearest lepton or jet.  Compared to \met,
\metrel\ has increased rejection power when the \met\ is generated by
a neutrino in a jet or the mis-measurement of an object.

The data are subdivided into \ZeroJet, \OneJet{} and \TwoJet{} search
channels since the background rate and composition depend
significantly on the jet multiplicity. The \ZeroJet{} background is
dominated by $WW$ events, and top is an important background for the
other two channels. To reduce the $WW$ background, the di-leptons are
required to be close together ($\left|\Delta\phi_{\ell\ell}\right| <
1.8$), which arises from the spin-0 of a SM Higgs and the V-A nature
of the $W$ decay. Top backgrounds are reduced with $b$-tagging
requirements.  Further details on the event selection can be found
in~\cite{ATLAS:2012af}.  For \mH\,=\,125\,GeV, the combined acceptance
times efficiency of the 8\,TeV \ZeroJet{} and \OneJet{} selection is
about 7.4\%.

\subsection{Background normalisation and control samples}
\label{sec:control}

The leading backgrounds from SM processes producing two isolated
high-$\pt$ leptons and $\MET$ are $WW$ and top, both $t\bar{t}$ and
single top.  These are estimated using partially data-driven
techniques based on normalising the MC predictions to the data in
control regions dominated by the relevant background source.  The
\Wjets{} background is estimated from data for all jet multiplicities.
Only the small backgrounds from Drell-Yan and di-boson processes other
than $WW$, as well as the $WW$ background for the \TwoJet{} analysis,
are estimated using MC simulation.

The control regions are defined by selections similar to those used
for the signal region but with some criteria reversed or modified to
obtain signal-depleted samples enriched in a particular background.
Some control regions have significant contributions from backgrounds
other than the targeted one, which introduces dependencies among the
background estimates.  These correlations are taken into account in
the $WW$ control region where the top and \Wjets{} backgrounds are
subtracted using their respective measurements.
See~\cite{ATLAS:2012af} for full details on the background control
samples and estimates.

\subsection{Systematic uncertainties}
\label{sec:systematics}

The systematic uncertainties that have the largest impact on the
sensitivity of the search are the theoretical uncertainties associated
with the signal, arising from the separation into 0-,1-,and 2-jet
channels.  The main experimental uncertainties are associated with the
JES, the jet energy resolution (JER), pile-up, \met{}, the $b$-tagging
efficiency, the $W$+jets transfer factor, and the integrated
luminosity.  The largest uncertainties on the backgrounds include $WW$
normalisation and modelling, top normalisation, and $W\gamma^{(\ast)}$
normalisation. The \TwoJet{} systematic uncertainties are dominated by
the statistical uncertainties in the data and the MC simulation.


\begin{table}[!htbp]
  \centering
  \caption{
    The expected numbers of signal (\mH\,=\,125\,GeV) and background
    events after all selections, including a cut on the transverse
    mass of $0.75\,m_{H}<m_{\rm T}<m_{H}$ for \mH\,=\,125\,GeV.  The
    observed numbers of events in data are also displayed.  The
    uncertainties shown are the combination of the statistical and all
    systematic uncertainties, taking into account the constraints from
    control samples.  For the \TwoJet\ analysis, backgrounds  with
    fewer than 0.01 expected events are marked with `-'.
    Ref.~\protect\cite{ATLAS:2012af}.   
  }
  \label{hww_cutflow}
  \vspace*{0.2cm}
  \scalebox{0.8}{
    \begin{tabular}{l|r@{$\,\pm \,$}l r@{$\,\pm \,$}l r@{$\,\pm \,$}l}
      \hline
      \hline
      & \multicolumn{2}{c}{\ZeroJet} & \multicolumn{2}{c}{\OneJet} &  \multicolumn{2}{c}{\TwoJet}  \\
      \hline
      Signal & 20 & 4 & 5 & 2 & 0.34 & 0.07 \\ \hline
      $WW$ & 101 & 13 & 12 & 5 & 0.10 & 0.14 \\ 
      $WZ^{(\ast)}/ZZ/W\gamma^{(\ast)}$ & 12 & 3 & 1.9 & 1.1 & 0.10 & 0.10 \\ 
      $t\bar{t}$ & 8 & 2 & 6 & 2 & 0.15 & 0.10 \\ 
      $tW/tb/tqb$  & 3.4 & 1.5 & 3.7 & 1.6 & \multicolumn{2}{c}{-} \\ 
      $Z/\gamma^{\ast}+\mathrm{jets}$ & 1.9 & 1.3 & 0.10 & 0.10 &\multicolumn{2}{c}{-} \\ 
      $W+\mathrm{jets}$ & 15 & 7 & 2 & 1 & \multicolumn{2}{c}{-} \\ \hline
      Total Background & 142 & 16 & 26 & 6 & 0.35 & 0.18 \\ \hline
      Observed  & \multicolumn{2}{c}{185} & \multicolumn{2}{c}{38} &  \multicolumn{2}{c}{0}  \\
      \hline
      \hline
    \end{tabular}
  }
\end{table}

\subsection{Results}
\label{sec:hwwresults}

Table~\ref{hww_cutflow} shows the numbers of events expected from a SM
Higgs boson with \mH\,=\,125\,GeV and from the backgrounds, as well
as the numbers of candidates observed in data, after application of
all selection criteria plus an additional cut on $\mT$ of $0.75\,m_{H}
< m_{\rm T} < m_{H}$.  The uncertainties shown in
Table~\ref{hww_cutflow} combine statistical and systematic
uncertainties.  An excess of events relative to the background
expectation is observed in the data.

\begin{figure}[hbtp]
  \centering
  \includegraphics[width=0.60\textwidth]{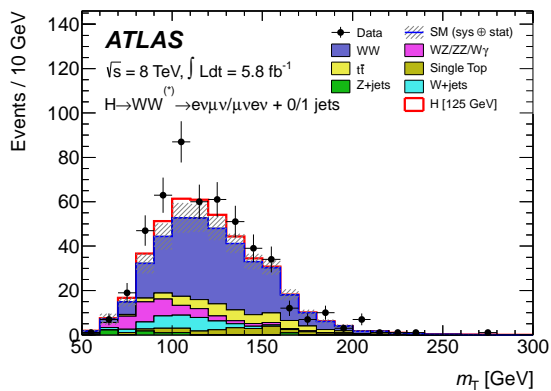}
  \vspace*{-0.5cm}
  \caption{ Distribution of the transverse mass, \mT, in the
    \ZeroJet{} and \OneJet{} analyses, for events satisfying all
    selection criteria.  The expected signal for \mH\,=\,125\,GeV is
    shown stacked on top of the background prediction.  The $W$+jets
    background is estimated from data, and $WW$ and top background MC
    predictions are normalised to the data using control regions.  The
    hashed area indicates the total uncertainty on the background
    prediction. Ref.~\protect\cite{ATLAS:2012af}.  }
  \label{fig:papermT}
\end{figure}

Figure~\ref{fig:papermT} shows the distribution of the transverse mass
after all selection criteria in the \ZeroJet{} and \OneJet{} channels
combined.  The statistical interpretation of the observed excess of
events is presented in Section~\ref{sec:Results}.

\section{Statistical procedure\label{sec:statproc}}

The parameter of interest is the global signal strength factor $\mu$,
which is a scale factor on the total number of events predicted for a
SM Higgs boson signal, defined such that $\mu=0$ corresponds to the
background-only hypothesis and $\mu=1$ corresponds to the SM Higgs
boson signal plus background.  Hypothesised values of $\mu$ are tested
with a statistic $\lambda(\mu)$ based on the profile likelihood
ratio~\cite{Cowan:2010st}.  This test statistic extracts the
information on the signal strength from a full likelihood fit to the
data, including all systematic uncertainties and their correlations.

Exclusion limits are based on the $CL_s$
prescription~\cite{Read:2002hq}; a value of $\mu$ is excluded at
95\%~CL when $CL_s$ is less than 5\%.  A SM Higgs boson mass $\mh$ is
excluded at 95\% CL when $\mu=1$ is excluded at that mass.  The
significance of an excess in the data is first quantified with the
local $p_0$, the probability that the background can produce a
fluctuation greater than the excess observed in data.  The equivalent
formulation in terms of number of standard deviations, $Z_l$, is
referred to as the local significance.  The global probability for the
most significant excess to be observed anywhere in a given search
region includes a correction for the ``look elsewhere'' effect,
reducing that given by local $p_0$ .

\section{Correlated systematic uncertainties}
\label{sec:CombSyst}

The full list of the individual search channels that enter the
combination are provided in~\cite{ATLAS:2012af}.  The main
uncorrelated systematic uncertainties correspond to the elements of
the background estimates.  The sources of correlated systematic
uncertainties are: integrated luminosity, electron/photon energy
scales, muon reconstruction, JES and $\MET$, and theoretical
uncertainties.

\section{Results\label{sec:Results}}

The addition of the 8\,TeV data for the \hZZllll, \hgg\ and
$\hwwenmun$ channels, as well as the improvements to the analyses of
the 7\,TeV data in the first two of these channels, bring a
significant gain in sensitivity in the low-mass region with respect to
the previous combined search~\cite{paper2012prd}.

\subsection{Excluded mass regions}

\begin{figure}[!htb]
  \begin{center}
    \subfigure[\label{fig:CLlimit}]{\includegraphics[width=0.49\linewidth]{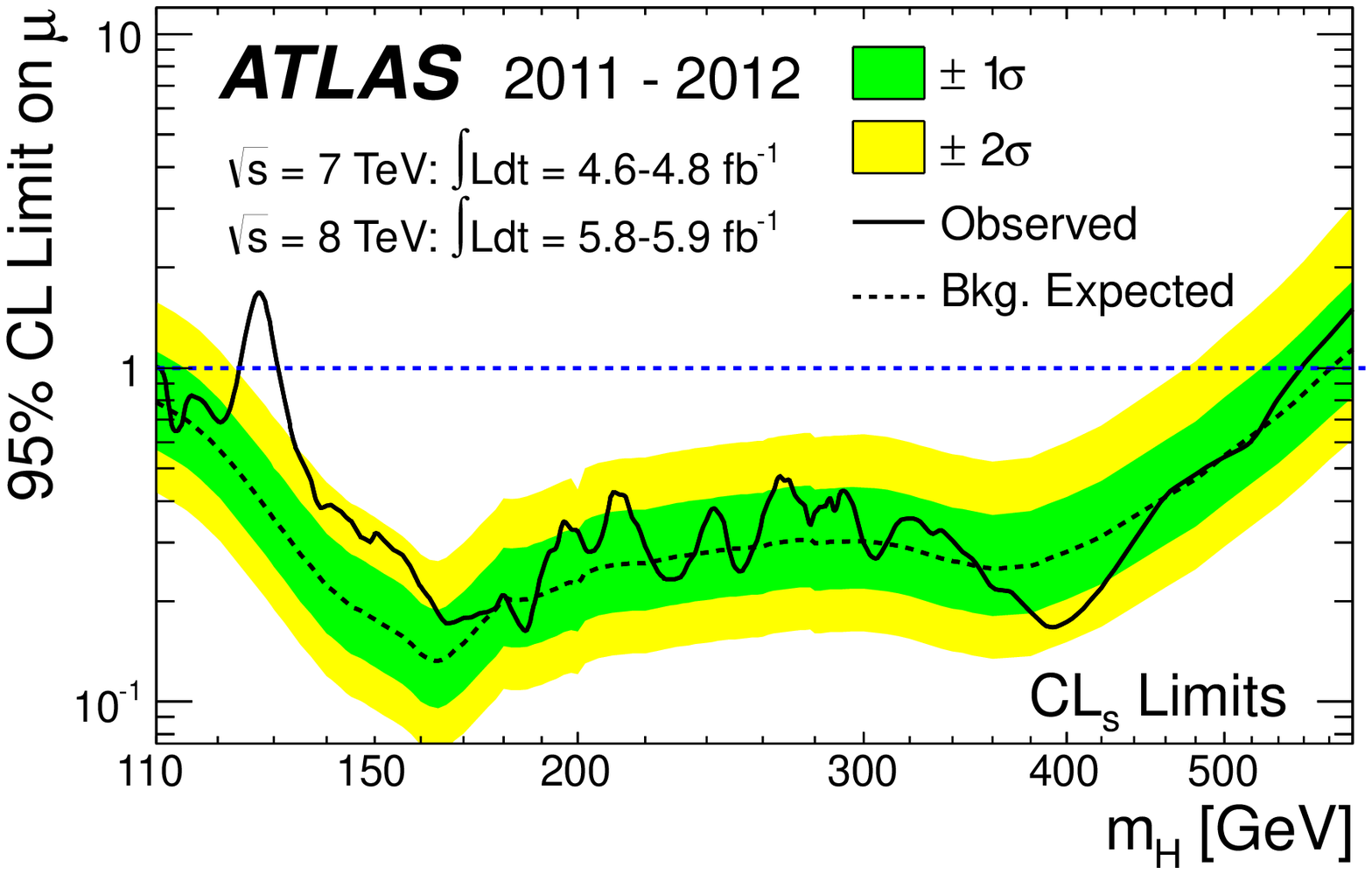}}
    \subfigure[\label{fig:p0}]{\includegraphics[width=0.49\linewidth]{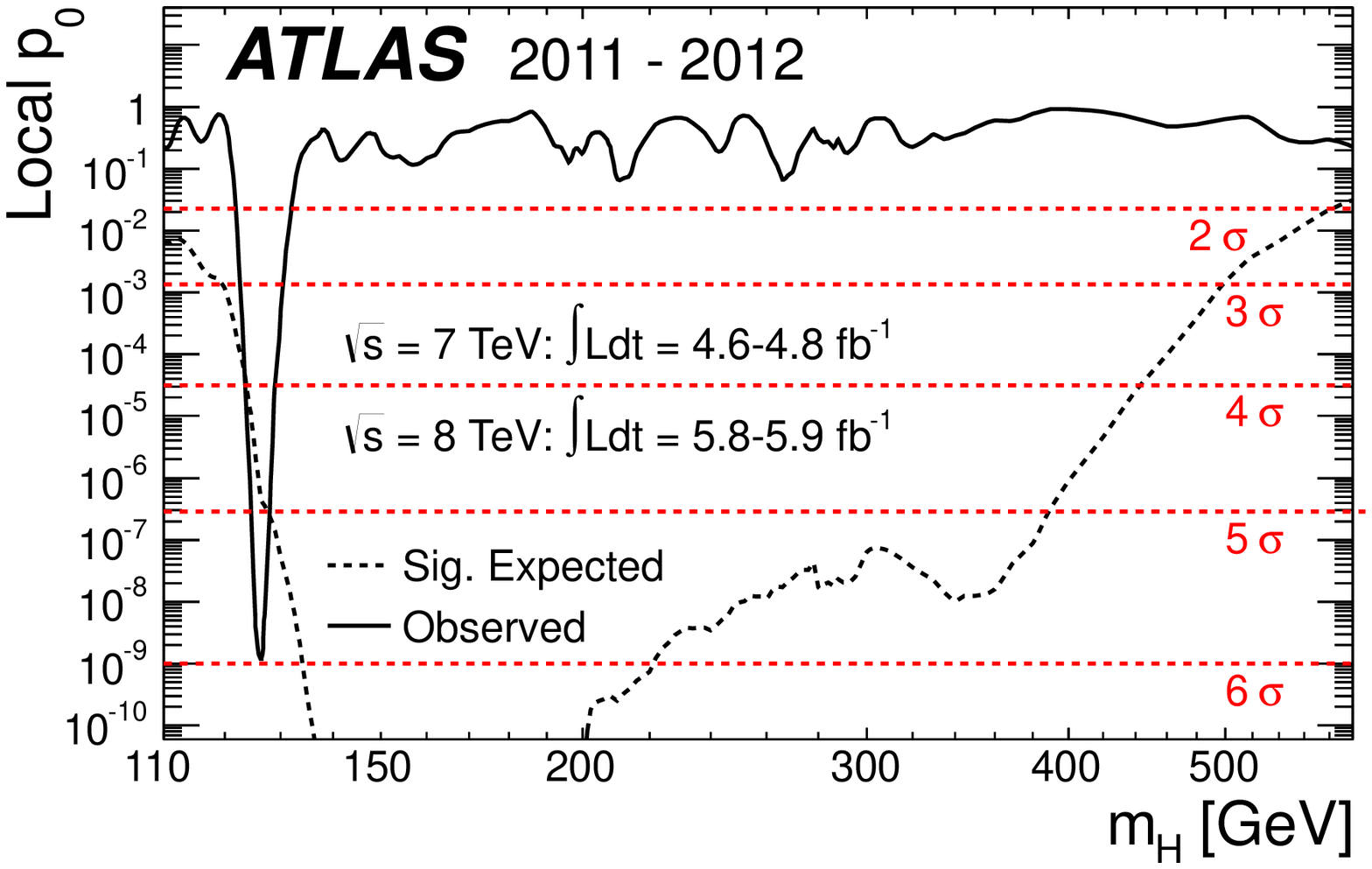}}
    \subfigure[\label{fig:mu}]{\includegraphics[width=0.49\linewidth]{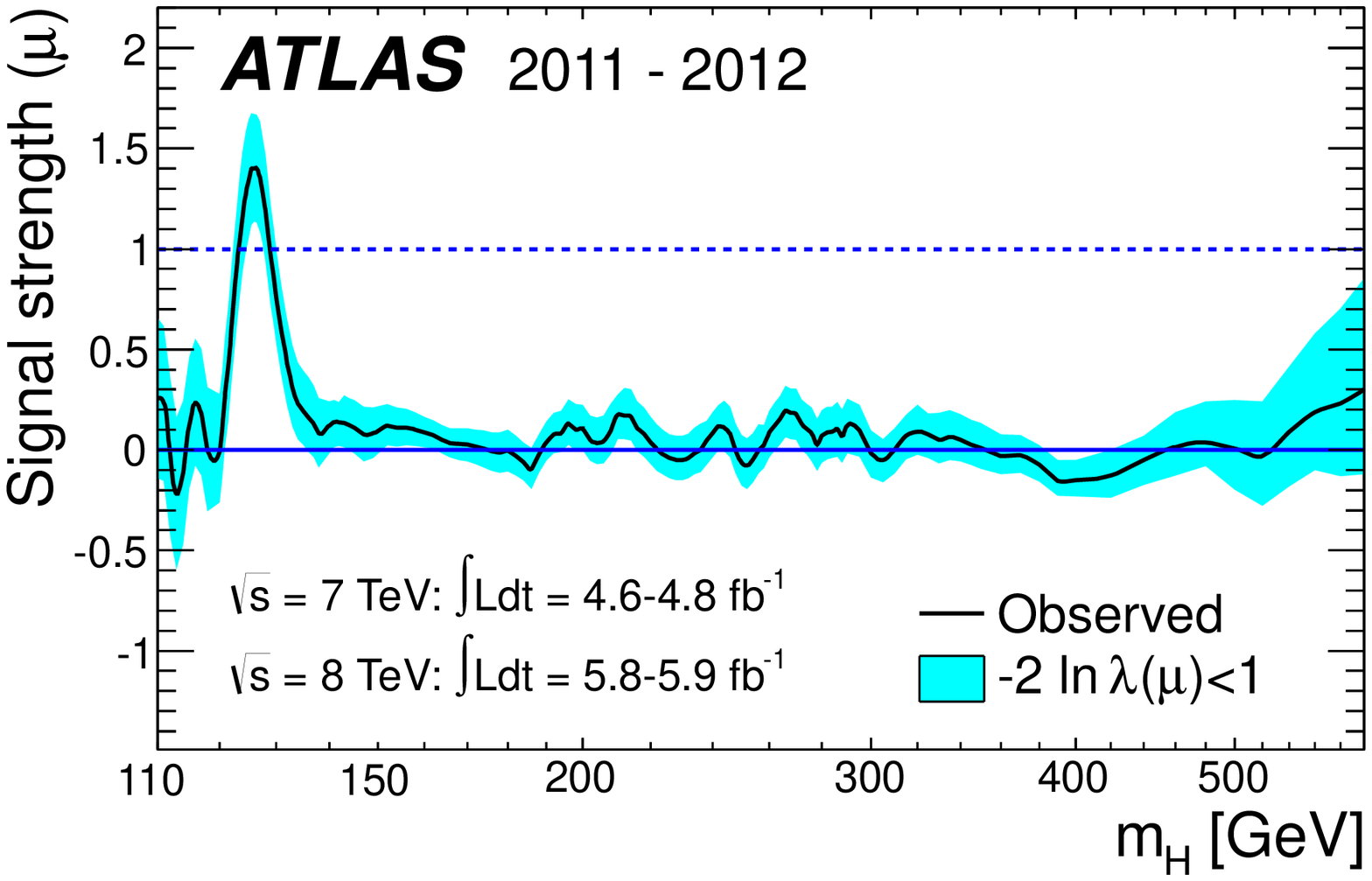}}
    \vspace*{-0.5cm}
    \caption{Combined search results: (a) The observed (solid) 95\%~CL
      limits on the signal strength as a function of \mh\ and the
      expectation (dashed) under the background-only hypothesis. The
      dark and light shaded bands show the $\pm 1\,\sigma$ and $\pm
      2\,\sigma$ uncertainties on the background-only expectation. (b)
      The observed (solid) local $p_0$ as a function of \mh\ and the
      expectation (dashed) for a SM Higgs boson signal hypothesis
      ($\mu=1$) at the given mass. (c) The best-fit signal strength
      $\hat\mu$ as a function of \mh. The band indicates the
      approximate 68\%\ CL interval around the fitted
      value. Ref.~\protect\cite{ATLAS:2012af}. } 
    \label{fig:CLsetal}
  \end{center}
\end{figure}

The combined 95\%~CL exclusion limits on the production of the SM
Higgs boson, expressed in terms of the signal strength parameter
$\mu$, are shown in Fig.~\ref{fig:CLsetal}(a) as a function of \mh.
The expected 95\%~CL exclusion region covers the \mH\ range from
\lowerExp\ to \upperExp. The observed 95\%~CL exclusion regions are
\lowerlowerObsNoGeV --\upperlowerObs\ and \lowerObsNoGeV --\upperObs.
Three mass regions are excluded at 99\%~CL, 113--114, 117--121 and
132--527\,GeV, while the expected exclusion range at 99\%\,CL is
113--532\,GeV.


\begin{figure}[!htb]
  \begin{center}
    \subfigure[\label{fig:p0zz}]{\includegraphics[width=0.49\linewidth]{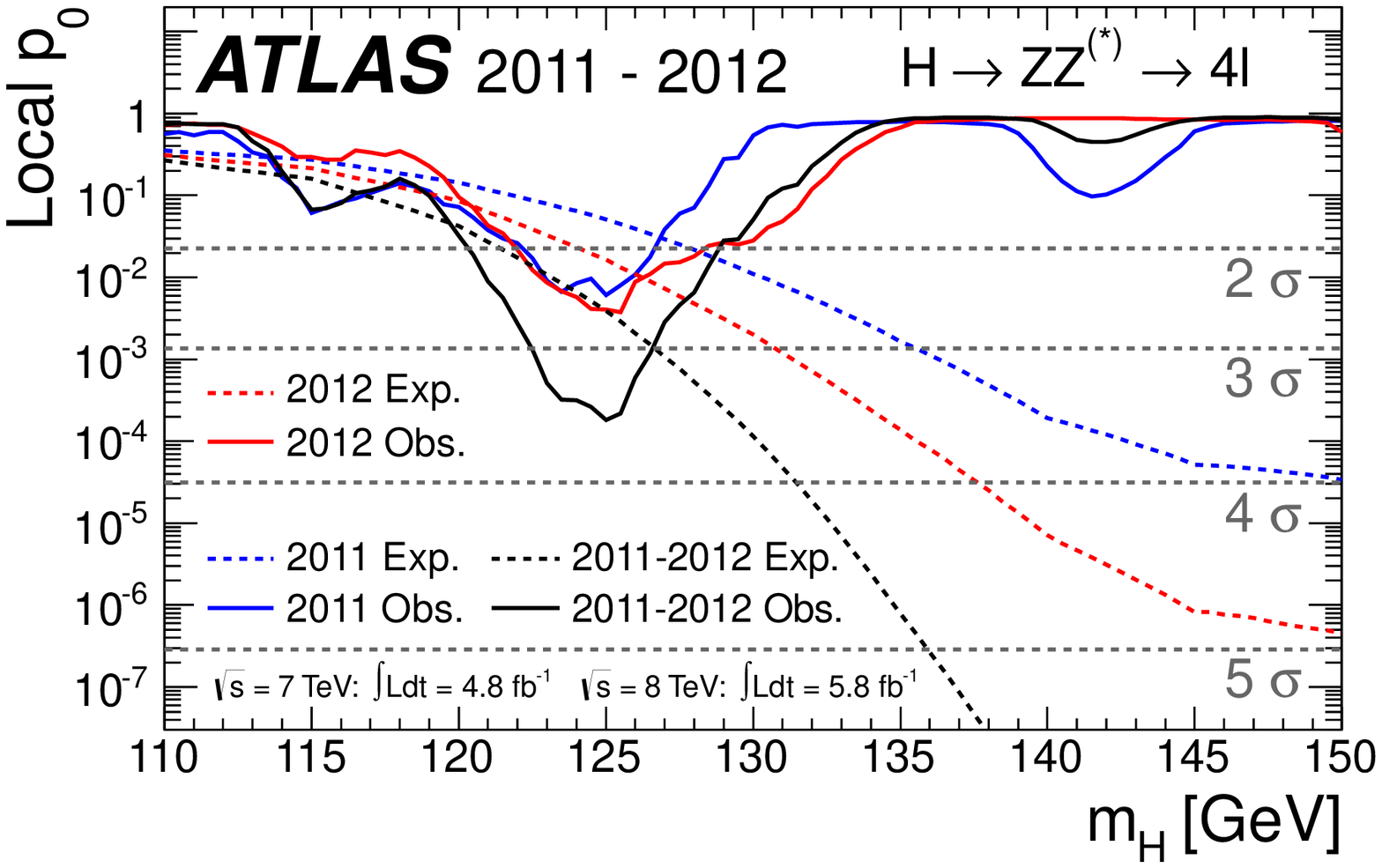}}
    \subfigure[\label{fig:p0gg}]{\includegraphics[width=0.49\linewidth]{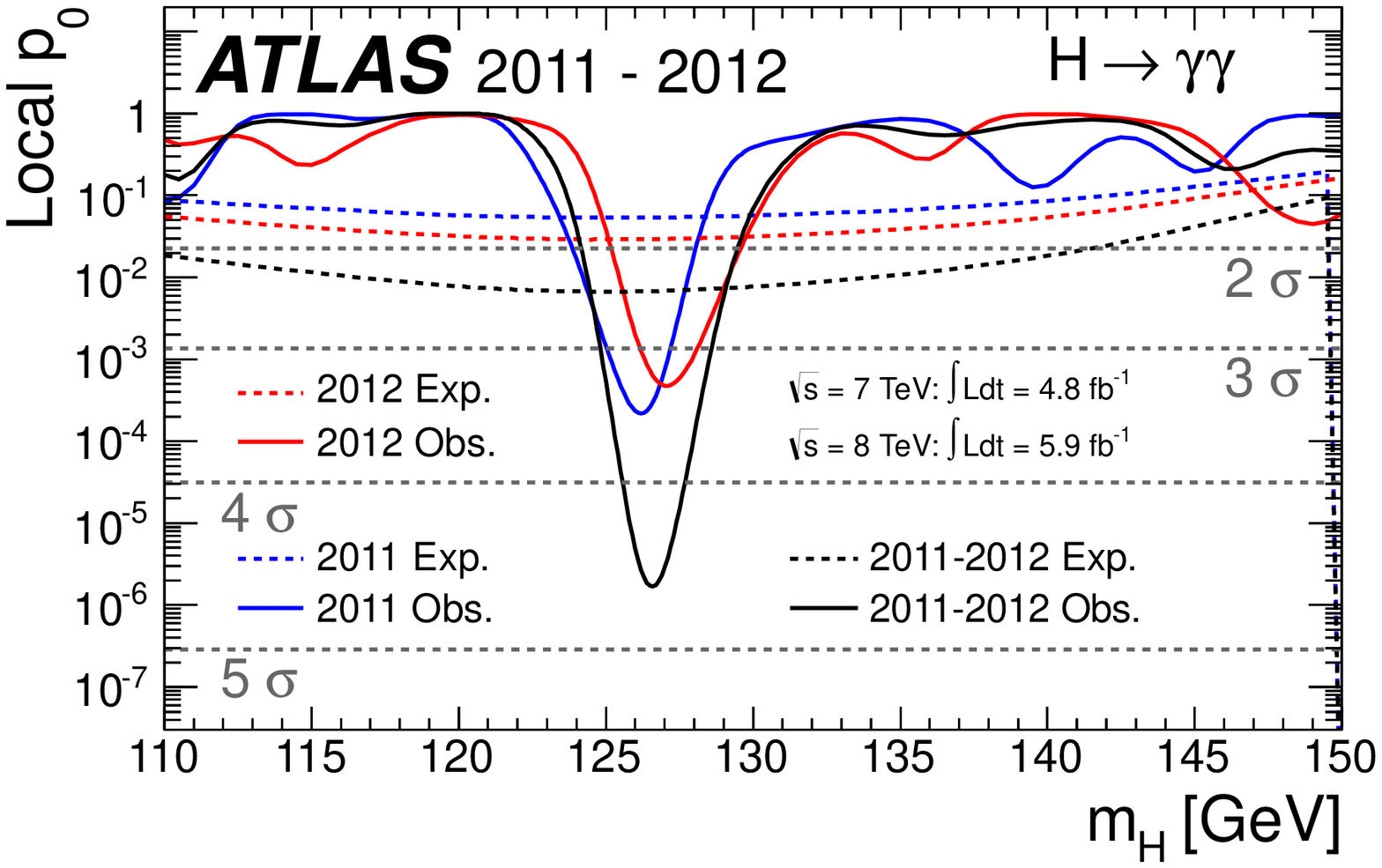}}
    \subfigure[\label{fig:p0ww}]{\includegraphics[width=0.49\linewidth]{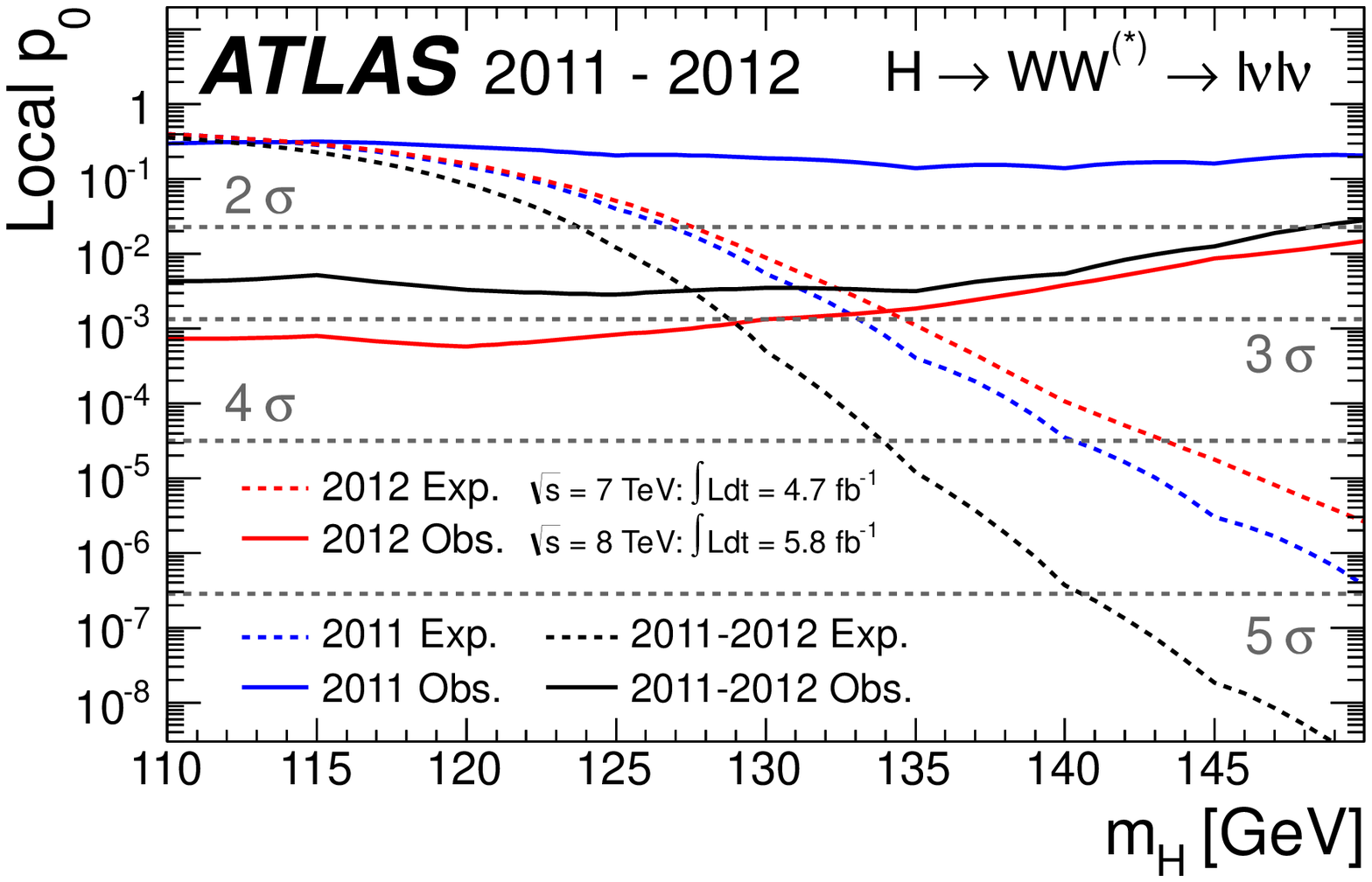}}
    \vspace*{-0.5cm}
    \caption{The observed local $p_0$ as a function of the
      hypothesised Higgs boson mass for the (a) \htollll, (b) \hgg\
      and (c) \hwwlnln\ channels.  The dashed curves show the expected
      local $p_0$ under the hypothesis of a SM Higgs boson signal at
      that mass.  Results are shown separately for the
      $7$\,TeV data (dark blue), the $8$\,TeV data
      (light red), and their combination
      (black). Ref.~\protect\cite{ATLAS:2012af}. } 
    \label{fig:allp0}
  \end{center}
\end{figure}

\begin{table*}[!htb]
  \centering
  \caption{Characterisation of the excess in the \htollllp, \hgg\ and
    \hwwlnln\ channels and the combination of all channels.  The mass
    value $m_{\rm max}$ for which the local significance is maximum,
    the maximum observed local significance $Z_l$ and the expected
    local significance $E(Z_l)$ in the presence of a SM Higgs boson
    signal at $m_{max}$ are given. The best fit value of the signal
    strength parameter $\hat\mu$ at \mh\,=\,126\,GeV is shown with the
    total uncertainty.  The expected and observed mass ranges excluded
    at 95\%\ CL (99\%\ CL, indicated by a~*) are also given, for the
    combined 7\,TeV and 8\,TeV data.
    Ref.~\protect\cite{ATLAS:2012af}. \label{tab:statsummary}}   
  \vspace{0.3cm}
  \resizebox{\textwidth}{!}{
    \begin{tabular}{cc|ccc|c|cc}
      \hline\hline
      Search channel & Dataset & $m_{\rm max}$ [GeV] &  $Z_l\,[\sigma]$   & $E(Z_l)\,[\sigma]$ & $\hat{\mu}(\mh=126$\,GeV) & Expected exclusion [GeV] & Observed exclusion [GeV] \\
      \hline
      \multirow{3}{*}{\htollllp} & 7\,TeV   & 125.0   & 2.5 & 1.6 & $1.4\pm 1.1$  \\
      & 8\,TeV   & 125.5   & 2.6 & 2.1 & $1.1\pm 0.8$  \\
      & 7 \& 8\,TeV & 125.0& 3.6 & 2.7 & $1.2\pm 0.6$& \excludedrangeexpaBrief,  \excludedrangeexpbBrief & \excludedrangeaBrief, \excludedrangebBrief \\
      \hline
      \multirow{3}{*}{\hgg}  & 7\,TeV      & 126.0      & 3.4 & 1.6  & $2.2\pm 0.7$\\
      & 8\,TeV      & 127.0      & 3.2 & 1.9  & $1.5\pm 0.6$     \\
      & 7 \& 8\,TeV & 126.5      & 4.5 & 2.5  & $1.8\pm 0.5$& 110--140 &  112--123, 132--143  \\
      \hline
      \multirow{3}{*}{\hwwlnln} & 7\,TeV      & 135.0  & 1.1 & 3.4 & $0.5\pm 0.6$\\
      & 8\,TeV      & 120.0  & 3.3 & 1.0 & $1.9\pm 0.7$\\
      & 7 \& 8\,TeV & 125.0  & 2.8 & 2.3 & $1.3\pm0.5$& 124--233 & 137--261 \\
      \hline
      \multirow{4}{*}{Combined}   & 7\,TeV      & 126.5  & 3.6 & 3.2 & $1.2\pm 0.4$&         &   \\
      & 8\,TeV      & 126.5  & 4.9 & 3.8 & $1.5\pm 0.4$&         &   \\ 
      & \multirow{2}{*}{7 \& 8\,TeV} & \multirow{2}{*}{126.5}  & \multirow{2}{*}{6.0} &\multirow{2}{*}{4.9} & \multirow{2}{*}{$1.4\pm 0.3$} & \lowerExpNoGeV--\upperExpNoGeV & \lowerlowerObsNoGeV--\upperlowerObsNoGeV, \lowerObsNoGeV--\upperObsNoGeV \\
      &                              &                         &                      & &                               &                   113--532 (*) & 113--114, 117--121, 132--527 (*) \\
      \hline\hline
    \end{tabular}} 
\end{table*}

\subsection{Observation of an excess of events}

An excess of events is observed near \mh\,=\,126\,GeV in the 
\hZZllll\ and \hgg\ channels, both of which provide fully reconstructed
candidates with high resolution in invariant mass, as shown in
Figures~\ref{fig:allp0}(a) and \ref{fig:allp0}(b). These excesses
are confirmed by the highly sensitive but low-resolution 
$\hwwlnln$ channel, as shown in Fig.~\ref{fig:allp0}(c).

\begin{figure}[!htb]
  \centering
  \subfigure[\label{fig:CLball}]{\includegraphics[width=0.49\linewidth]{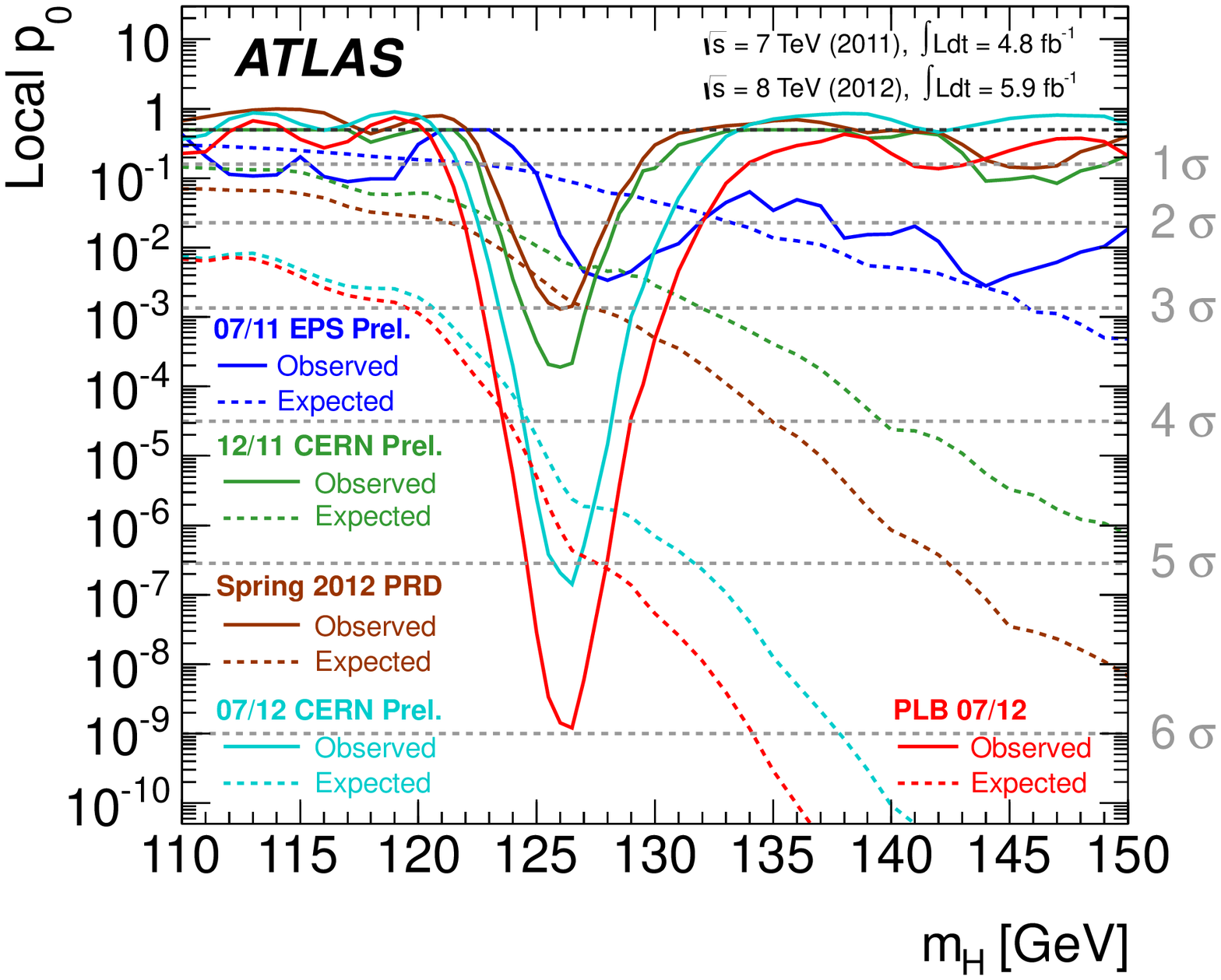}}
  \subfigure[\label{fig:mubarchart}]{\includegraphics[width=0.44\linewidth]{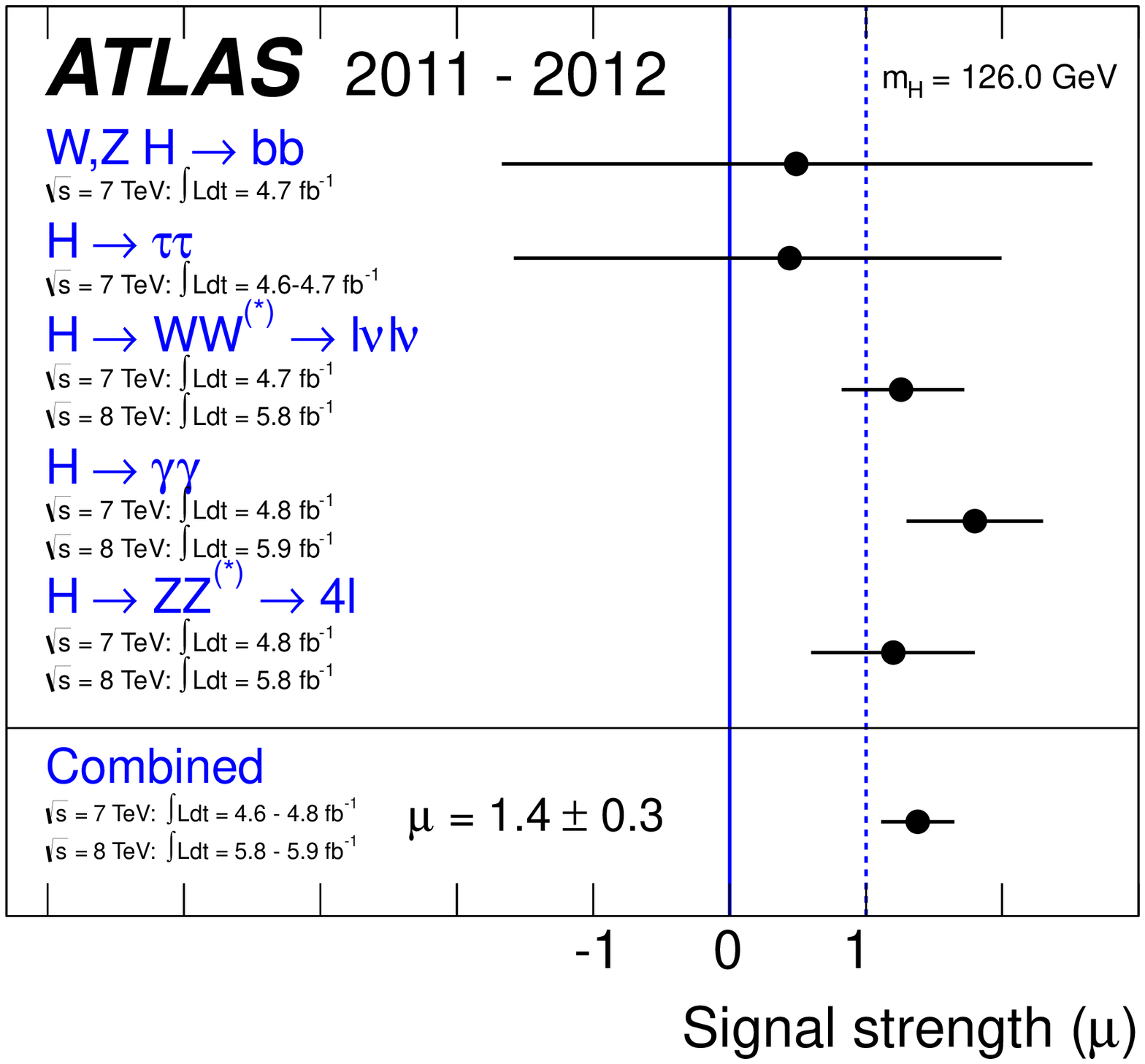}}
  \vspace*{-0.5cm}
  \caption{\subref{fig:CLball} The observed (solid) local $p_0$ as a
    function of \mh\ in the low mass range. The dashed curve shows the
    expected local $p_0$ under the hypothesis of a SM Higgs boson
    signal at that mass with its $\pm 1\,\sigma$ band. Results are
    shown as a function of time.  The different sets of lines The
    horizontal dashed lines indicate the $p$-values corresponding to
    significances of 1 to $6\,\sigma$.  \subref{fig:mubarchart}
    Measurements of the signal strength parameter $\mu$ for
    \mh\,=\,126\,GeV for the individual channels and their
    combination.  Ref.~\protect\cite{ATLAS:2012af}. }\label{fig:CLb}
\end{figure}

The observed local $p_0$ values from the combination of channels,
using the asymptotic approximation, are shown as a function of \mh\ in
Fig.~\ref{fig:CLsetal}(b) for the full mass range and in
Fig.~\ref{fig:CLball} for the low mass range, shown as a function
of time.

The largest local significance for the combination of the 7 and 8\,TeV
data is found for a SM Higgs boson mass hypothesis of
\mH\,=\,126.5\,GeV, where it reaches \significance, with an expected
value at that mass of \expectedsignificance\ (see also
Table~\ref{tab:statsummary}).  For the 2012 data alone, the maximum
local significance for the \hZZllll, \hgg\ and $\hwwenmun$ channels
combined is $4.9\,\sigma$, and occurs at \mH\,=\,126.5\,GeV
($3.8\,\sigma$ expected). Including electron/photon energy resolution
and scale systematic uncertainties, as described
in~\cite{ATLAS-CONF-2012-093}, reduces the local significance to
$5.9\,\sigma$.  The global significance of a local \significanceESS\
excess anywhere in the mass range 110--600\,GeV is estimated to be
approximately $5.1\,\sigma$.

\subsection{Characterising the excess}

The mass of the observed new particle is estimated using the profile
likelihood ratio $\lambda(m_H)$ for \hZZllll\ and \hgg, the two
channels with the highest mass resolution.  The signal strength is
allowed to vary independently in the two channels, although the result
is essentially unchanged when restricted to the SM hypothesis $\mu=1$.
The leading sources of systematic uncertainty come from
the electron and photon energy scales and resolutions.
The resulting estimate for the mass of the observed particle is
\massresultStatSys.

The best-fit signal strength $\hat{\mu}$ is shown in
Fig.~\ref{fig:CLsetal}(c) as a function of \mh.  The observed excess
corresponds to $\hat{\mu} = 1.4\pm 0.3$ for \mH\,=\,126\,GeV,
consistent with the SM Higgs boson hypothesis $\mu=1$. A summary of
the individual and combined best-fit values of the strength parameter
for a SM Higgs boson mass hypothesis of 126\,GeV is shown in
Fig.~\ref{fig:mubarchart}, while more information about the three main
channels is provided in Table~\ref{tab:statsummary}.

\begin{figure}[!hbt]
  \center
  \subfigure[\label{fig:ggFvsVBFcoup}]{\includegraphics[width=0.49\linewidth]{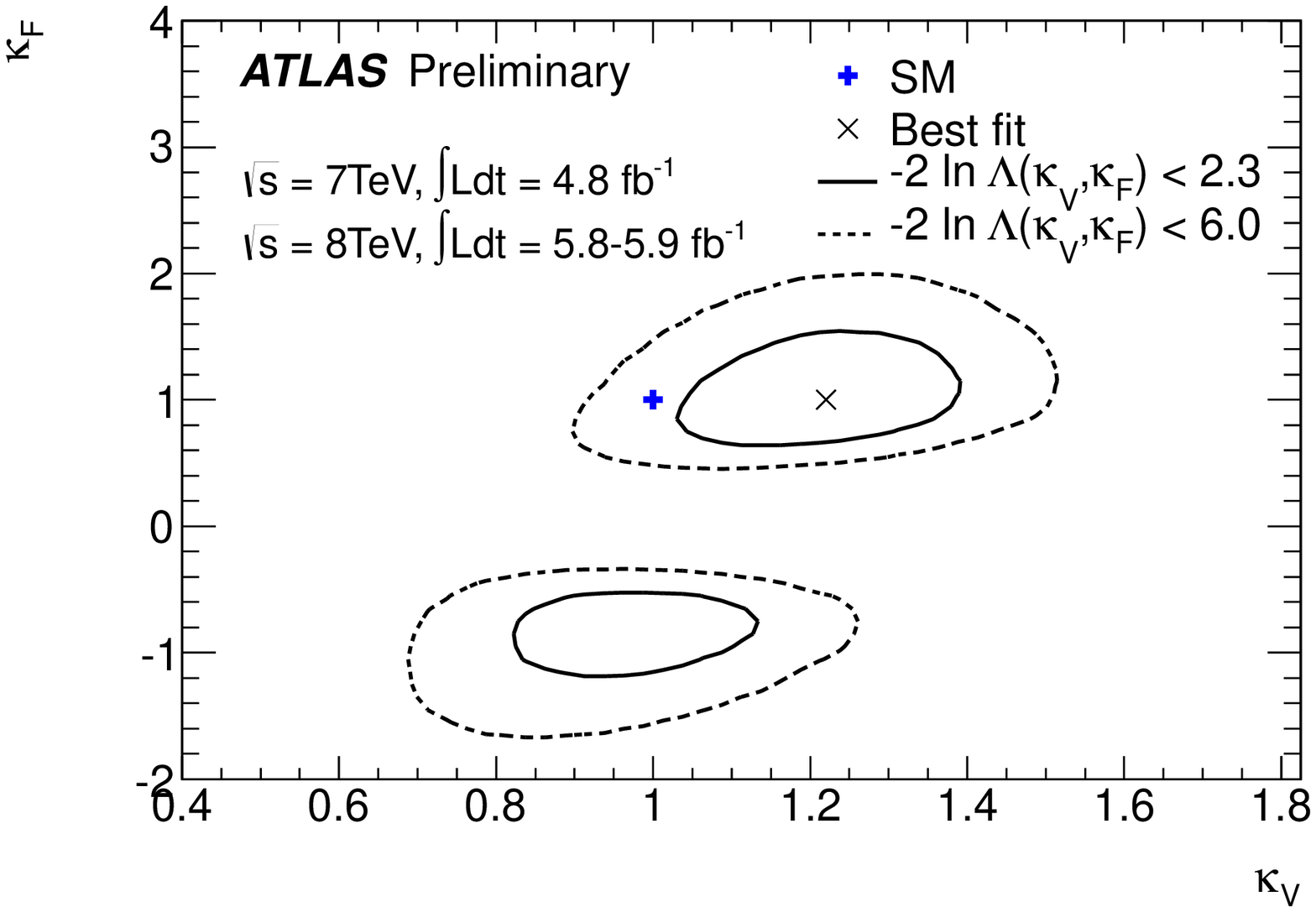}}
  \subfigure[\label{fig:WZcoup}]{\includegraphics[width=0.49\linewidth]{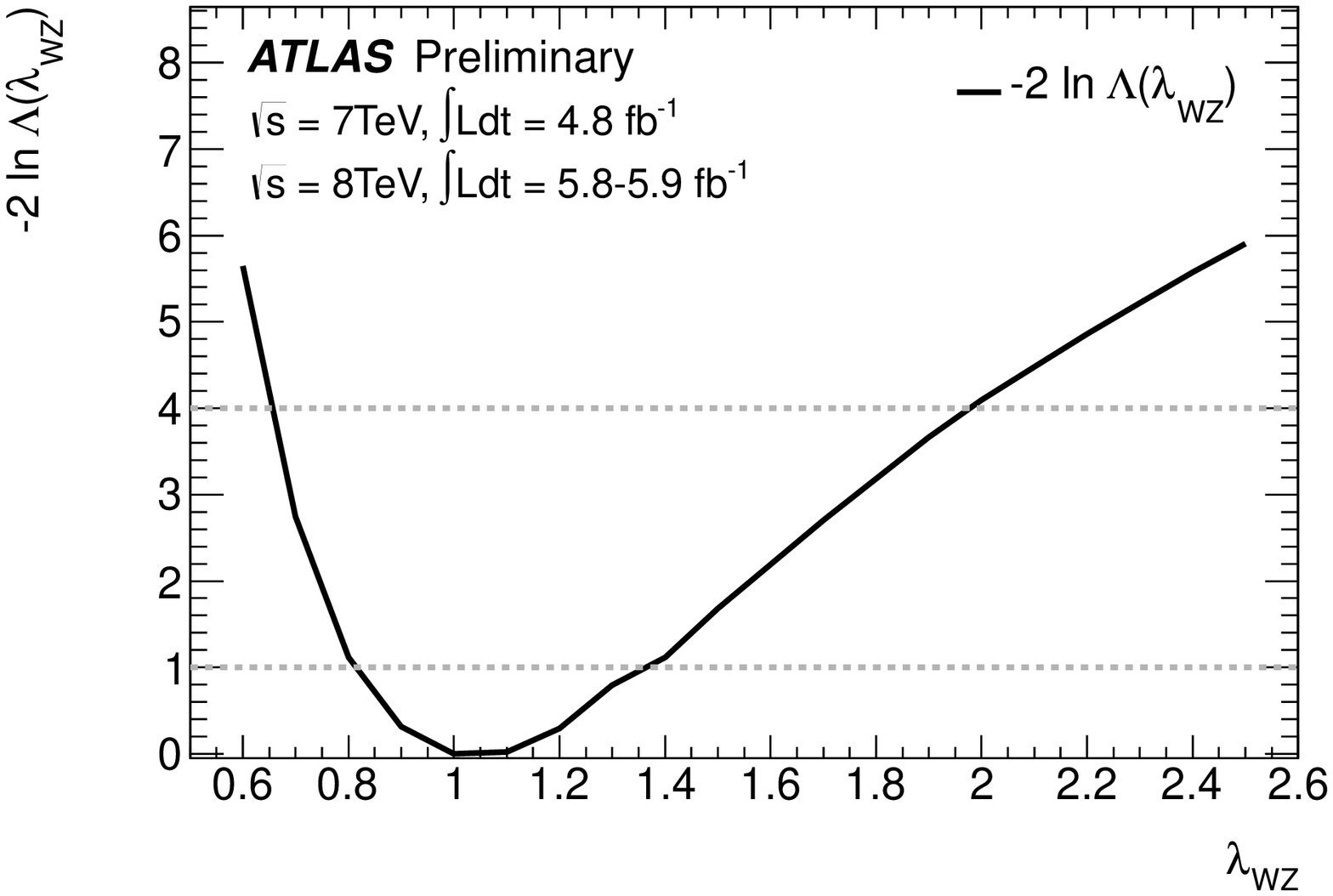}}
  \subfigure[\label{fig:invisCoup}]{\includegraphics[width=0.49\linewidth]{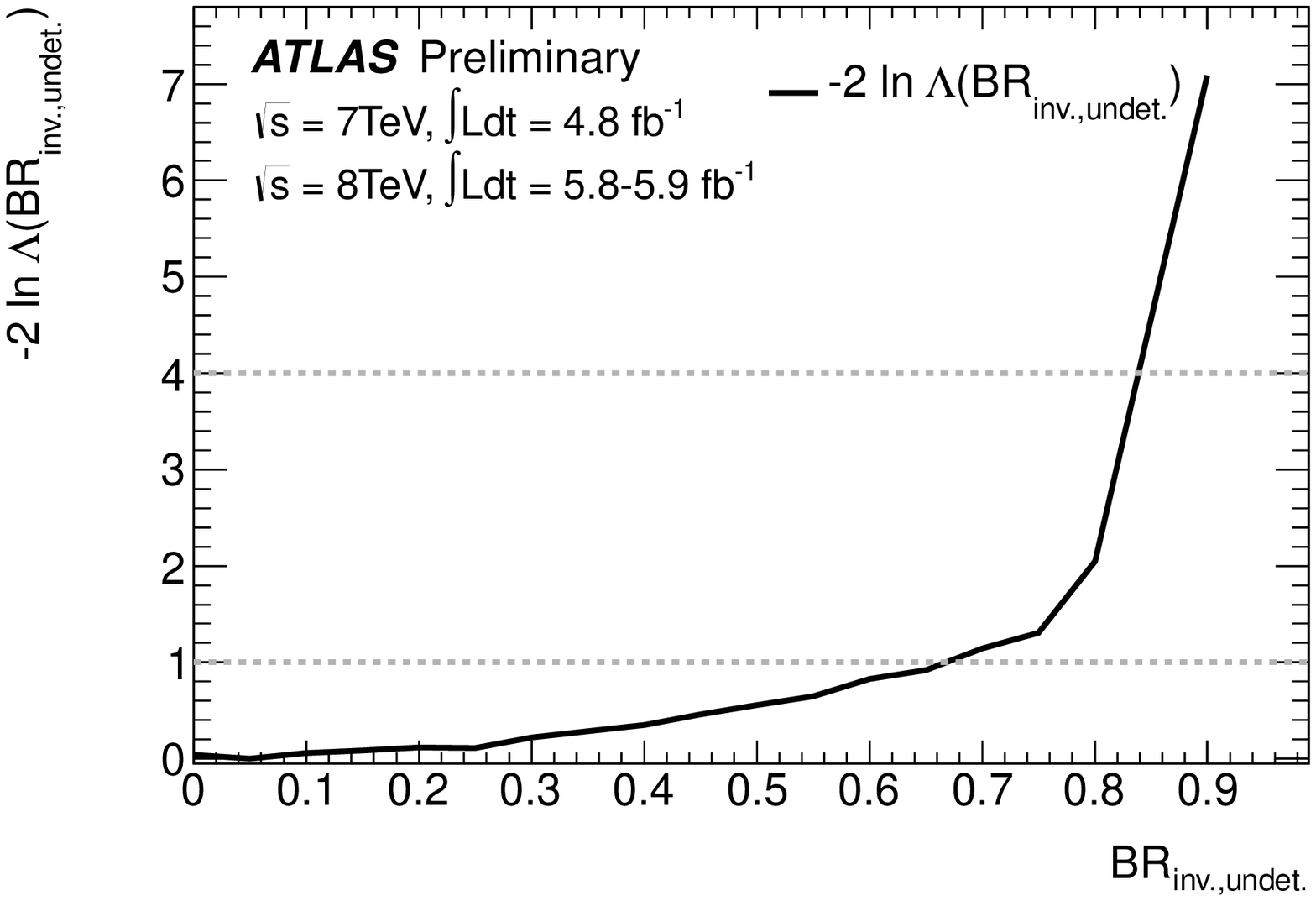}}
  \vspace*{-0.5cm}
  \caption{\subref{fig:ggFvsVBFcoup} Fit for different coupling
    strengths for fermions and vector bosons, assuming no non-SM
    contribution to the total width; \subref{fig:WZcoup} fit probing
    deviations in the vector sector; \subref{fig:invisCoup} fit of
    invisible or undetectable branching ratio when probing $gg \to H$
    and $\hgg$ loops. The contours and horizontal dashed lines
    correspond to the $68\%$ and $95\%$
    CL. Ref.~\protect\cite{CouplingsConf}. } 
  \label{fig:couplings}
\end{figure}

The observed production and decays modes allow the SM Higgs couplings
to be probed~\cite{CouplingsConf} with a few simple assumptions: a
single resonance with \mh\,=\,126\,GeV, SM Higgs $J^{CP}$ $(0^{++})$,
and a negligible width (i.e.\ $\sigma \times BR(ii\to H \to ff) =
\sigma_{ii} \times \Gamma_{ff}/\Gamma_H$ for initial/final states
$i/f$). Compatibility of with a SM Higgs interpretation can be
expressed by scale factors $\Cc_i$ such that $\sigma_{ii}/\sigma_{SM}
= \Gamma_{ii}/\Gamma_{SM} = \Cc^2_i$. Figure~\ref{fig:couplings} shows
the $\Cc_i$ likelihood fit results for three simplifying assumptions.
Fig.~\ref{fig:ggFvsVBFcoup} shows $\Cc_F$ vs $\Cc_V$ assuming a single
scale factor for all fermions and for all vector couplings. The $68\%$
CL intervals when profiling over all other parameters are: $\Cc_F \in
[-1.0,0.7] \cup [0.7,1.3]$, $\Cc_V \in [0.9,1.0] \cup [1.1,1.3]$.
Fig.~\ref{fig:WZcoup} shows the ratio of $W$ to $Z$ couplings
$\lambda_{WZ} = \Cc_W/\Cc_Z$. These scale factors are required to be
identical within tight bounds by $\rm SU(2)_V$ custodial symmetry and
the $\rho$ parameter measurements at LEP~\cite{lepew:2010vi}. The
fitted ratio is: $\lambda_{WZ} = 1.07^{+0.35}_{-0.27}$. Finally,
Fig.~\ref{fig:invisCoup} shows an invisible or undetectable branching
ratio when allowing $\Cc_g$ and $\Cc_\gamma$ to vary, providing at
$68\%$ CL $\rm BR_{inv.,undet.} < 0.68$. No significant deviation from
SM expectations is found.

\section{Conclusion\label{sec:Conclusion}}

Searches for the SM Higgs boson have been performed in the \htollll,
\hgg\ and \hWWenmun\ channels with the ATLAS experiment at the LHC
using 5.8--5.9~\infb\ of $pp$ collision data recorded in 2012 at
8\,TeV. These results are combined with earlier
results~\cite{paper2012prd}, which are based on an integrated
luminosity of 4.6--4.8~\infb\ recorded in 2011 at 7\,TeV, except for
the \htollll\ and \hgg\ channels, which have been updated with the
improved analyses.

The SM Higgs boson is excluded at 95\%\ CL in the mass range
\lowerlowerObsNoGeV --\upperObs, except for the narrow region
\upperlowerObsNoGeV --\lowerObs. In this region, an excess of events
with a local significance \significanceESS, corresponding to
$p_0=1.7\times 10^{-9}$, is observed. The excess is driven by the two
channels with the highest mass resolution, \htollll\ and \hgg, and the
equally sensitive but low-resolution \hWWlnln\ channel.  Taking into
account the entire mass range of the search, 110--600\,GeV, the global
significance of the excess is $5.1\,\sigma$, which corresponds to
$p_0=1.7\times 10^{-7}$.

These results provide conclusive evidence for the discovery of a new
particle with mass \massresultStatSys. The signal strength parameter
$\mu$ has the value $1.4\pm 0.3$ at the fitted mass, which is
consistent with the SM Higgs boson hypothesis $\mu=1$.  The decays to
pairs of vector bosons whose net electric charge is zero identify the
new particle as a neutral boson.  The observation in the di-photon
channel disfavours the spin-1 hypothesis~\cite{landau1948,yang1950}.
Tests of the production and decay couplings with simplifying
assumptions find that these results are compatible with the hypothesis
that the new particle is the SM Higgs boson, and more data are needed
to assess its nature in detail.


\bibliographystyle{atlasBibStyleWithTitle} 
\providecommand{\href}[2]{#2}\begingroup\raggedright

\end{document}